%% file: paper.tex
\documentclass[11pt,a4paper]{article}
\pdfoutput=1   
\usepackage{multirow}
\usepackage{multicol}
\usepackage{alpha}

\usepackage{macros}
\usebiblio{mainbib}
\begin{document}

\input{title.tex}

\input{s1.tex}

\input{s2.tex}

\input{s3.tex}

\input{s4.tex}

\begin{acknowledgement}%
We like to thank R.~Sommer for many helpful discussions and contributions to
the study presented here, and we thank our colleagues in the CLS effort for the
joint production and use of gauge configurations.
We acknowledge partial support by the SFB/TR~9, by grant HE~4517/2-1 (P.F. and
J.H.) and HE~4517/3-1 (J.H.) of the Deutsche Forschungsgemeinschaft, by
the European Community through EU Contract MRTN-CT-2006-035482, ``FLAVIAnet'',
by the Spanish Minister of Education and Science projects RyC-2011-08557 and 
by the Danish National Research Foundation under the grant n. DNRF:90 (M.D.M.).
P.F. acknowledges financial support from the Spanish MINECO under grants
FPA2012-31880 and SEV-2012-0249 (``Centro de Excelencia Severo Ochoa''
Programme). N.G is supported by the Leverhulme trust, research grant RPG-2014-118. 

We gratefully acknowledge the computer resources granted by the John von
Neumann Institute for Computing (NIC) and provided on the supercomputer JUROPA
at J\"ulich Supercomputing Centre (JSC) and by the Gauss Centre for
Supercomputing (GCS) through the NIC on the GCS share of the supercomputer
JUQUEEN at JSC, with funding by the German Federal Ministry of Education and
Research (BMBF) and the German State Ministries for Research of
Baden-W\"urttemberg (MWK), Bayern (StMWFK) and Nordrhein-Westfalen (MIWF), as
well as within the Distributed European Computing Initiative by the PRACE-2IP,
with funding from the European Community's Seventh Framework Programme
(FP7/2007-2013) under grant agreement RI-283493, by the Grand \'Equipement
National de Calcul Intensif at CINES in Montpellier under the allocation
2012-056808, by the HLRN in Berlin, and by NIC at DESY, Zeuthen.
\end{acknowledgement}

\begin{appendix}
  \input{app_ana.tex}
  \input{app_Eeff.tex}

\end{appendix}

\end{document}

%% file: title.tex
\preprintno{%
CP3-Origins-2015-009 DNRF90,
DESY 15-062,\\
DIAS-2015-9,
HU-EP-15/16,
IFT-UAM/CSIC-15-043,\\
LPT-Orsay/15-31,
MITP/15-025,
MS-TP-15-07,
TCDMATH 15-04\\[3em]
}

\title{%
B-meson spectroscopy in HQET at order $1/m$
}

\collaboration{\includegraphics[width=2.8cm]{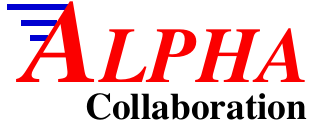}}

\author[desy,dresden]{Fabio~Bernardoni}
\author[fra]{Beno\^it~Blossier}
\author[trin]{John~Bulava}
\author[ode,esp]{Michele~Della~Morte}
\author[hu,uam]{Patrick~Fritzsch}
\author[ply]{Nicolas~Garron}
\author[fra,fra2]{Antoine~G\'erardin}
\author[wwu]{Jochen~Heitger}
\author[mainz]{Georg~von~Hippel}
\author[desy]{Hubert~Simma}

\address[desy]{John von Neumann Institut for Computing (NIC), DESY, Platanenallee~6, 15738~Zeuthen, Germany}
\address[dresden]{Medizinische Fakult\"at, Carl Gustav Carus, TU Dresden, Fetscherstra{\ss}e~74, 01307~Dresden, Germany}
\address[fra]{Laboratoire~de~Physique~Th\'eorique, Universit\'e~Paris~XI,  91405~Orsay~Cedex, France}
\address[fra2]{Laboratoire~de~Physique~Corpusculaire, Universit\'e~Blaise~Pascal, \\CNRS/IN2P3 63177~Aubi\`ere~Cedex, France}
\address[ply]{School of Computing and Mathematics, Plymouth~University, PL4~8AA~Plymouth, UK}
\address[trin]{School~of~Mathematics, Trinity~College, Dublin~2, Ireland}
\address[ode]{CP$^3$-Origins \& Danish IAS, University of Southern Denmark, Campusvej 55, 5230 Odense M, Denmark}
\address[esp]{IFIC and CSIC, Calle Catedr\'atico Jos\'e Beltran 2, 46980~Paterna, Valencia, Spain}
\address[hu]{Institut~f\"ur~Physik, Humboldt-Universit\"at~zu~Berlin, Newtonstr.~15, 12489~Berlin, Germany}
\address[uam]{Instituto~de~F\'isica~Te\'orica~UAM/CSIC, Universidad Aut\'onoma de Madrid, C/~Nicol\'as~Cabrera~13-15, Cantoblanco E-28049~Madrid, Spain}
\address[wwu]{Institut~f\"ur~Theoretische~Physik, Universit\"at~M\"unster, Wilhelm-Klemm-Str.~9, 48149~M\"unster, Germany}
\address[mainz]{Institut~f{\"u}r~Kernphysik, University~of~Mainz, Becherweg~45, 55099~Mainz, Germany}

\begin{abstract}
We present a study of the B spectrum performed in the framework of Heavy Quark
Effective Theory expanded to next-to-leading order in $\minv$ and
non-perturbative in the strong coupling. Our analyses have been performed on
$\Nf=2$ lattice gauge field ensembles corresponding to three different lattice
spacings and a wide range of pion masses.  We obtain the $\Bs$-meson mass and
hyperfine splittings of the B- and $\Bs$-mesons that are in good agreement with
the experimental values and examine the mass difference $\mBs-\mB$ as a further
cross-check of our previous estimate of the b-quark mass. We also report on the
mass splitting between the first excited state and the ground state in the B
and $\Bs$ systems.
\end{abstract}


\maketitle

%% file: s1.tex
\section{Introduction}\label{sec:intro}

Lattice studies of the hadron spectrum provide valuable indications on the reliability
of effective field theories.  A phenomenologically relevant example is given by
effective field theories for heavy quarks, such as Heavy Quark Effective Theory
(HQET) or Non-Relativistic QCD (NRQCD).
In this respect, hyperfine splittings are of great theoretical interest, since
they vanish in the static ($\minvh \to \infty$) limit, and thus serve to probe
subleading terms in the heavy quark expansion.  

Regarding specifically the hyperfine splitting of the B spectrum, it is well 
known that the masses of the pseudoscalar $\Bq$ and vector  $\Bstarq$ differ 
as a result of spin effects
\cite{PDG}.
This mass difference is produced in HQET by a $\Or(\minvh)$ chromomagnetic
term in the Lagrangian that breaks the heavy quark spin symmetry of the
static theory
\cite{Eichten:1990vp,Neubert:1993mb}.
In the context of lattice NRQCD it was found that in order to obtain a
determination of the hyperfine splittings of bottomonium that is consistent
with experiments, the perturbative improvement of the NRQCD action to at least
one-loop order is required
\cite{Hammant:2011bt,Hammant:2013sca,Dowdall:2013jqa}, for values of the
lattice spacing of about 0.1 fm.
The $\Upsilon$ spectrum, on the other hand, is well reproduced within NRQCD,
and has been used in the literature as a quantity to set the lattice spacing
\cite{Davies:2009tsa,Dowdall:2011wh}.
Hyperfine splittings have also been studied within the HISQ approach to heavy 
quarks~\cite{Dowdall:2012ab}.

In contrast to other studies, our results are based on non-perturbatively 
renormalized HQET including the next-to-leading order terms in the inverse heavy
quark mass expansion. When formulated with a lattice regulator, the resulting lattice
field theory can be non-perturbatively matched to QCD in the continuum limit.
This provides a rigorous and systematically improvable approach to the study of
the B-meson system.

Although permil precision seems difficult to reach within this approach, as it
would require including $1/m_{\rm h}^2$ corrections, the control over
systematics such as the non-perturbative subtraction of power-divergences and
the related existence of the continuum limit, makes the approach very appealing
and conceptually sound and in this sense, and in our view, more precise than
other methods.
In addition, as evident in~\cite{Aoki:2013ldr} and emphasized
in~\cite{DellaMorte:2015rua}, for quantities more ``complicated'' than decay
constants, such as form-factors, there is great need for results from different
approaches, with different levels of control on the various systematics.
Applications of the non-perturbative HQET formalism to form-factors are well on
their way~\cite{DellaMorte:2015yda,DellaMorte:2013ega,Bahr:2014iqa}.

Remaining within spectrum-related observables, another quantity of interest is
the SU(3) isospin breaking difference $\mBs-\mB$.  Most of the dependence on
the heavy quark mass cancels in the difference and the statistical correlations
between the strange and light-quark measurements should lead to a much improved
precision in the determination of the difference compared to what would be
possible for the individual masses. 
In
\cite{Bernardoni:2013xba},
the mass of the B-meson has been used to determine the b-quark mass.
Since the mass of the $\Bs$-meson could have equally well been used,
it is instructive to study how the strange-light mass difference propagates
to the b-quark mass measurement.

Relatively little is known experimentally on the radial excitations of
B-mesons, usually denoted by $\Bprime$. CDF has claimed the observation of a
resonant state ${\rm B}(5970)$ that may be interpreted as a $\Bprime$ state
\cite{Aaltonen:2013atp}.
Lattice predictions for the radial excitation energies of the B system have
been made in the framework of NRQCD
\cite{Collins:1999ff, Gregory:2010gm}
and HQET
\cite{Michael:1998sg, Green:2003zza, Burch:2006mb, Foley:2007ui,
      Koponen:2007nr, Koponen:2007fe, Burch:2008qx, Jansen:2008si,
      Michael:2010aa},
but, with the exception of the most recent HQET studies, control of the continuum
extrapolation has been limited, if possible at all.
An additional difficulty that always needs to be addressed is in disentangling
single particle excitations from multi-particle states. We use the notation
$\Br$ (or $\Brq$ with $q={\rm u}/{\rm d}$ or ${\rm s}$) for the excited states 
that we observe. They might be identified with radial excitations $\Bprime$, 
but without a dedicated study including also multi-hadron operators, we
are unable to conclusively determine the nature of these excited states.

In this paper, we report on our estimate of the hyperfine splitting 
$\mBstar-\mB$ in the B and $\Bs$ systems, and on the mass differences
$\mBr-\mB$, $\mBrs- \mBs$ and $\mBs-\mB$ from $\Nf=2$ lattice simulations.

%% file: s2.tex
\section{Theoretical setup}\label{sec:setup}

\subsection{HQET on the lattice}\label{s:lathqet}

The HQET action at $\Or(\minvh)$ reads
\begin{align}         \label{e:hqetaction}
  \Shqet 
      &= a^4{\sum}_x \big\{ \Lstat(x) - \omega_{\rm kin} \Okin(x) - \omega_{\rm spin} \Ospin(x) \big\} \,,  \\ \label{e:lstat}
  \Lstat(x) 
      &= \heavyb(x)\,  D_0                  \,\heavy(x)\,, \\[0.2em] \label{e:ofirst}
  \Okin(x) 
      &= \heavyb(x)\, \vecD^2               \,\heavy(x)\,, \\[0.2em]
  \Ospin(x) 
      &= \heavyb(x)\, \vecsigma \cdot \vecB \,\heavy(x)\,,
\end{align}
where the subscript $\rm h$ denotes a heavy (static) quark field
satisfying $\frac{1+\gamma_0}{2}\heavy=\heavy$, and the operators are
normalized such that the classical (tree level) values of the coefficients are
$\omegakin=\omegaspin=1/(2\mh)$. The energy levels computed in this theory are
relative to a bare mass $\mhbare$, which at tree level is simply $\mh$, but has
to absorb a power-divergent shift at the quantum level, implying that it will
take a different value depending on whether the $\Or(\minvh)$ terms are
included or not.
\begin{table}[t]
  \small
  \centering
  \renewcommand{\arraystretch}{1.2}
  \input{tables/table_omega_zb.tex}

  \caption{Values of HQET parameters at the physical point
           $\boldsymbol{\omega}(z=\zb)$.
           As determined in~\cite{Bernardoni:2013xba},
           we have used $\zb^{\rm stat}=13.24$ to interpolate
           the parameters of HQET at static order, and $\zb=13.25$
           for the parameters of HQET expanded to NLO.
           The bare coupling $g_0$ is given by $\beta=6/g_0^2$.}
  \label{tab:omega_zb}
\end{table}
The renormalizability of the static theory can be preserved at NLO
by considering a strict expansion of correlation functions in $\minvh$,
\begin{equation}
  \langle O\rangle = \langle O\rangle_{\rm stat} +
    \omegakin \, a^4{\sum}_x \langle O\,\Okin(x) \rangle_\stat +
    \omegaspin\, a^4{\sum}_x \langle O\,\Ospin(x)\rangle_\stat  \;,
\end{equation}
where the suffix ``stat'' denotes an expectation value measured in the static
theory.

The values of the parameters $\mhbare^\stat$ and $\left\{\mhbare, \omegakin,
\omegaspin\right\}$ required to match to QCD have been determined in
\cite{Blossier:2012qu,Bernardoni:2013xba}
using the Schr\"odinger Functional scheme. In order to keep this paper
self-contained, we summarize in \tab{tab:omega_zb} their values for the three
lattice spacings $a(\beta)$ and two static discretizations (HYP1,
HYP2~\cite{DellaMorte:2005yc}) that we use.  The parameters are given
at $z_{\rm b}=L_1M_{\rm b}$ where $M_{\rm b}$ is the Renormalization Group Invariant
(RGI) b-quark mass and $L_1$ is the linear extent of the lattice used for the matching.

\subsection{The variational method}\label{s:gevp}

For spectroscopic applications, having good control over contributions from
excited states is crucial to reduce the systematic error on the determination
of the ground state.  For this purpose it is necessary to use a variational
method
\cite{Berg:1982kp,Griffiths:1983ah,Michael:1982gb,Michael:1985ne,Luscher:1990ck},
which starts from matrices of correlation functions,
\begin{eqnarray}  \nonumber
  C^{\rm{stat}}_{ij}(t) &=& \sum_{x,\bf{y}} \left< O_i(x_0+t,{\bf y})\,O^*_j(x)\right>_\stat\,,
  \\[-0.7em]            \label{e:cmatdefs}
  \\[-0.2em]            \nonumber
  C^{\rm{kin/spin}}_{ij}(t)& = & \sum_{x,\vecy,z}\;
  \left< O_i(x_0+t,\vecy)\,O^*_j(x)\,  {\cal O}_{\rm kin/spin}(z)\right>_\stat \,,
\end{eqnarray}
containing all pairwise correlators of some set of interpolating fields
$O_i\,,\;i=1,\ldots,N$, and proceeds to solve the generalized eigenvalue
problem (GEVP)
\begin{eqnarray} \label{e:gevp}
   C^{\rm stat}(t)\, v^{\rm stat}_n(t,t_0) = \lambda^{\rm stat}_n(t,t_0)\, C^{\rm stat}(t_0)\,v^{\rm stat}_n(t,t_0) \,,
   \quad n=1,\ldots,N\,,\quad t>t_0\;.
\end{eqnarray}
From the generalized eigenvalues $\lambda_n^\stat(t,t_0)$ and the corresponding
eigenvectors $v_n^\stat(t,t_0)$, the effective energy levels can be computed as
\cite{Luscher:1990ck,Blossier:2009kd}
\begin{align}\label{e:endef}
E_n^{\rm eff,\stat}(t,t_0) &= -{1\over a} \, [\log{\lambda_n^\stat(t+a,t_0)}-\log{\lambda_n^\stat(t,t_0)}] \,,\nonumber\\
E_n^{\rm eff,x}(t,t_0) &=  {\lambda_n^{\rm x}(t,t_0) \over 
                         \lambda_n^\stat(t,t_0)} \,-\,
      {\lambda_n^{\rm x}(t+a,t_0) \over 
                         \lambda_n^\stat(t+a,t_0)} \,,
\end{align}
where ${\rm x}\in\{{\rm kin},{\rm spin}\}$ and
\begin{align}\label{e:lamdef}
{\lambda_n^{\rm x}(t,t_0)\over \lambda_n^\stat(t,t_0)}
  &= \left(v_n^\stat(t,t_0)\,,\,
             [[\lambda_n^\stat(t,t_0)]^{-1}\,C^{\rm x}(t)- C^{\rm x}(t_0)] 
             v_n^\stat(t,t_0)\right)\,.
\end{align}
The corresponding asymptotic behaviour is known to be
\begin{align}
    E_n^{\rm eff,\stat}(t,t_0) &=
    E_n^\stat \,+\, \beta_n^\stat\,\rme^{-\Delta E_{N+1,n}^\stat\, t}+\ldots\,,
   \label{e:lamstatfit}
  \\
    E_n^{\rm eff,x}(t,t_0) &=
    E_n^{\rm x} \,+\, [\,\beta_n^{\rm x}
               \,-\, \beta_n^\stat\,t\,\Delta E_{N+1,n}^{\rm x}\,]
                     \rme^{-\Delta E_{N+1,n}^\stat\, t}+\ldots\, ,\qquad
   \label{e:lamfirstfit}
\end{align}
where the energy gap is defined as $\Delta E_{m,n}=E_m-E_n$.

In our study, the operator basis used is given by heavy-light bilinears
with varying levels of Gaussian smearing
\cite{Gusken:1989ad}
applied to the light quark field,
\begin{align}
 O_k(x) &= \heavyb(x)\,\gamma_0\gamma_5\,\psi_q^{(k)}(x)            \,,&
  \psi_q^{(k)}(x) &= \left( 1+\kappa_{\rm G}\,a^2\,\Delta \right)^{R_k} \psi_q(x) \,,
\end{align}
where $q={\rm u}/{\rm d}$ or ${\rm s}$. The covariant Laplacian $\Delta$ is
built from gauge links that have been triply APE smeared
\cite{smear:ape,Basak:2005gi}
in the spatial directions, and the smearing parameters $\kappa_{\rm G}$
and $R_k$ are chosen so as to approximately cover the same sequence of
physical radii at each value of the lattice spacing, as discussed in
\cite{Bernardoni:2013xba}.
We solve the GEVP for the matrix of correlators in the static limit for $N=3$. 

\begin{figure}[t]
  \centering
  \includegraphics[width=0.95\textwidth]{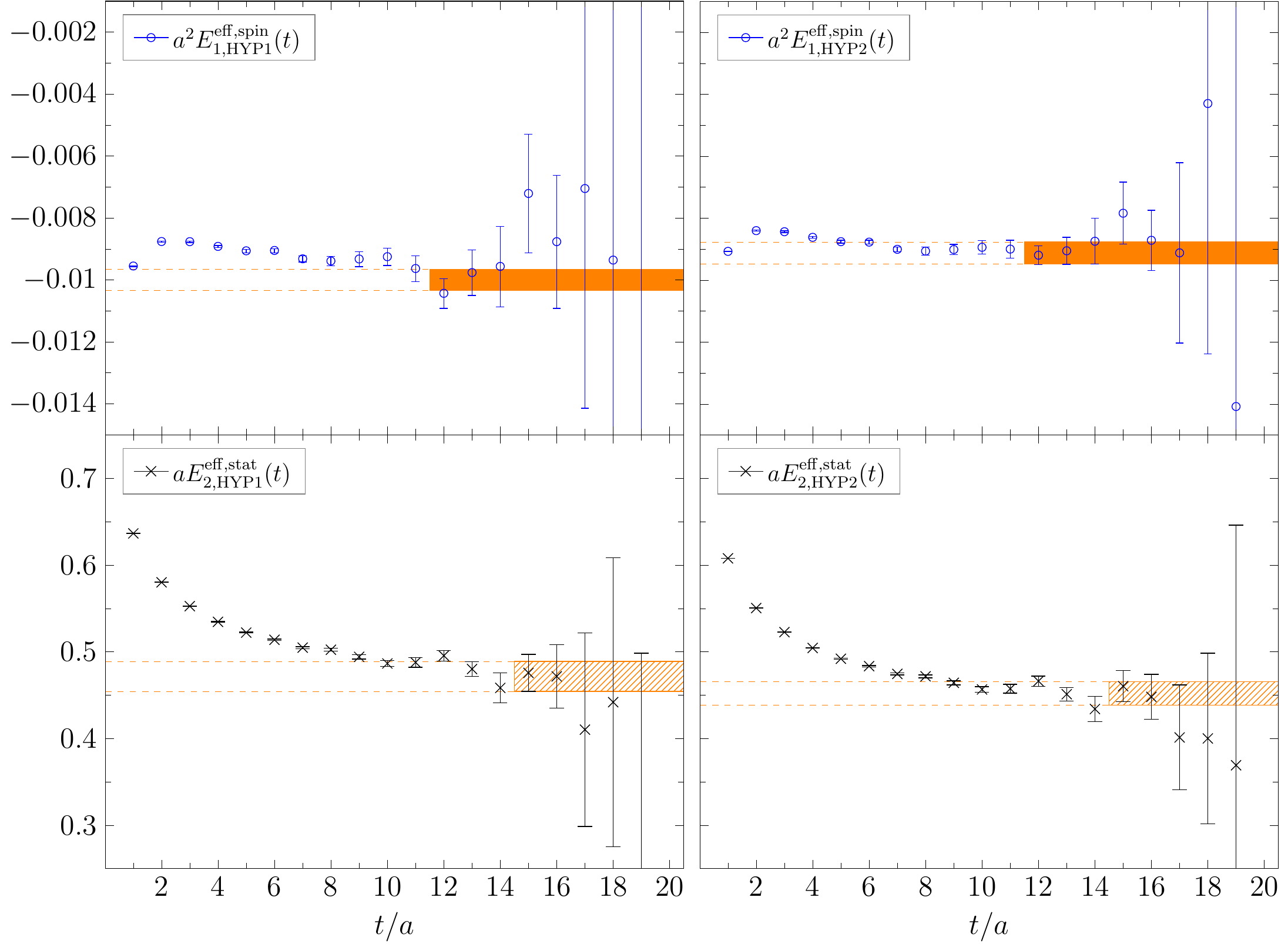}
  \vskip-1em
  \caption{Illustration of typical plateaux for the $\Or(\minv )$
           chromomagnetic energy $a^2E^{\rm spin}_{1}$ (\textit{top}) 
           and the first excited state static energy $aE^{\rm stat}_{2}$ 
           (\textit{bottom}), computed on the CLS ensemble N6 
           ($a=0.048$\,fm, $m_{\pi}=340$\,MeV).
          }
  \label{fig:plateaux}
\end{figure}

An illustration of two typical plateaux of the energies
$E_1^{\rm spin}$ and $E_2^{\rm stat}$ is shown in
Fig.~\ref{fig:plateaux}.
The plateau regions used for weighted averaging have been chosen by applying the
procedure already discussed in
\cite{Blossier:2010vz,Blossier:2010mk}
in order to ensure that the systematic errors due to excited-state
contributions are less than a given fraction (typically 1/3) of the statistical
errors on the GEVP results.  As a consistency check, we have also employed a
global fit of the form of eqs.~\eqref{e:lamstatfit} and \eqref{e:lamfirstfit}
to our data. The values of $E_n$ obtained from the fit are compatible with the
plateau values, and generally exhibit smaller statistical errors.  We therefore
consider our errors to be estimated conservatively.

\subsection{Computing masses}\label{s:obs}

We start by recalling the definition of the $\Bq$-meson mass as it has been
used to non-perturbatively fix the HQET parameters for the static quark
discretizations HYP1 and HYP2 in~\cite{Bernardoni:2013xba}. On the lattice one
combines the HQET parameters (see Table~\ref{tab:omega_zb},
or~\cite{Bernardoni:2014fva} for a more detailed discussion) with the large
volume computations of ground state energies, 
\begin{subequations}\label{eq:mBq}
\begin{align}\label{eq:mBq-stat}
    \mBq^{\rm stat}(\mpi,a) &= \mhbare^\stat + E_1^\stat\big|_{m_q} \,,\\
             \label{eq:mBq-HQET}
    \mBq           (\mpi,a) &= \mhbare + E_1^\stat\big|_{m_q} + \omegakin E_1^{\rm kin}\big|_{m_q} + \omegaspin E_1^{\rm spin}\big|_{m_q} \,.
\end{align}
\end{subequations}
We use the subscript $m_q$ to indicate that the corresponding energy results
from the GEVP analysis of correlation functions in the heavy-light ($q={\rm
d}$) or heavy-strange ($q={\rm s}$) sector along the lines presented in
Section~\ref{s:gevp}. The PDG value $\mB=5.2795\,\GeV$ in~\cite{PDG} 
has already been
used as input to determine $\mb$~\cite{Bernardoni:2013xba} and thus the HQET 
parameters $\omega_{i}$. However, re-computing $\mB$ at fixed values of the
HQET parameters serves as a non-trivial cross-check of our calculation. 
Additionally, we will compute $\mBs$ in the very same way. All remaining
observables presented in the present paper are mass splittings which can be
computed with the data resulting from our GEVP analysis. We now give their 
explicit definition before discussing any combined chiral and continuum
extrapolation in Section~\ref{sec:extrapolation}.

The hyperfine splitting of the $\Bq$-meson system, 
$\mBstarq-\mBq$, is given by
\begin{align}\label{eq:spinsplit}
       \dmHF{q}(\mpi,a)  &= - \frac{4}{3}\omegaspin E_1^{\rm spin} \big|_{m_q} \;, 
       &  q&= {\rm s, d} \;,
\end{align}
and depends on the HQET parameter $\omegaspin$ and the large-volume measurement
of $E_1^{\rm spin}$. The strange-light mass difference $\mBs-\mBd$,
at the static and $\Or(\minv)$ level, reads,
\begin{subequations}\label{eq:sl-split}
\begin{align}\label{eq:sl-split-stat}
    \dmijstat{\rm s-d}(\mpi,a) &= \Delta_{\rm s-d}E_1^\stat \,,\\
             \label{eq:sl-split-HQET}
    \dmij{\rm s-d}    (\mpi,a) &= \Delta_{\rm s-d}E_1^\stat + \omegakin \Delta_{\rm s-d}E_1^{\rm kin}
                                + \omegaspin \Delta_{\rm s-d}E_1^{\rm spin} \,,
\end{align}
\end{subequations}
respectively, where we have defined the shorthand
$ \Delta_{\rm s-d}Q = \left. Q \right|_{m_q=\mstrange} - \left. Q \right|_{m_q=\mdown}\,.$
Finally, the mass gaps for excited states, $\mBrq-\mBq$, can be computed from
$\Delta E_{m,n}=E_{m}-E_{n}$ via
\begin{subequations}\label{eq:radial}
\begin{align}  \label{eq:radialstat}
  \dmradstat{q}(\mpi,a)&= \Delta E_{2,1}^\stat\big|_{m_q} \,,\\  
               \label{eq:radialfirst}
  \dmrad{q}    (\mpi,a)&= \Delta E_{2,1}^\stat\big|_{m_q} 
             + \omegakin  \Delta E_{2,1}^{\rm kin} \big|_{m_q} 
             + \omegaspin \Delta E_{2,1}^{\rm spin}\big|_{m_q} \,,
\end{align}
\end{subequations}
to static and $\Or(\minv)$ accuracy, respectively. As any observable computed
on the lattice, these quantities depend on the input parameters used in the
simulation. The strange and bottom quark mass have been fixed to their physical
value along the lines reported
in~\cite{Fritzsch:2012wq,Bernardoni:2013xba,Bernardoni:2014fva}, such that only
the dependences on the light quark mass---here parameterized by $\mpi$---and
lattice spacing $a$ remain.

We have extracted the energies on a subset of
gauge ensembles from the Coordinated Lattice Simulations (CLS) effort for $\Nf=2$.
\begin{table}[tb]
  \centering\small
  \input{tables/table_measparms.tex}
  \caption{Measurement details for observables in the light- and strange-quark
           sector. For each ensemble we list the light- and strange-quark 
           hopping parameter, the trajectory length $\tau$, the distance of saved
           configurations $\tau_{\rm cfg}$ and the exponential autocorrelation
           time $\tau_{\rm exp}$ in MDU. 
           For each individual static quark discretization we then specify 
           $N_{\rm ms}$ [$m$,$d$], i.e., the total number of measurements starting
           from the $m^{\rm th}$ saved configuration using every $d^{\rm th}$.
           Note that for A5 we have two independent Monte-Carlo chains.
          }
  \label{tab:meas-parms}
\end{table}
The parameters of the simulations and the statistics entering our
analysis are collected in Table~1 of 
\cite{Bernardoni:2013xba}.
Remaining details of the measurements have been summarized in
Table~\ref{tab:meas-parms}.%
\footnote{%
For the B6-ensemble a preliminary value of $\kappa_{\rm s}$ has been used,
which produces a bare subtracted quark mass which differs by about 1 MeV from the
final one.
}
The light quark is treated in a unitary setup with
$\mpi$ in the range [$190\,\MeV$, $440\,\MeV$], and the bare strange quark mass
has been tuned on each CLS ensemble to its physical value
by using the kaon decay constant $\fK$ to set the scale
\cite{Fritzsch:2012wq}
and $\mK=494.2\,\MeV$.
The lattice spacings are $a/\fm\in\{0.048,0.065,0.075\}$ for
$\beta\in\{5.5,5.3,5.2\}$, corresponding to the CLS ensemble ids N--O, E--G and
A--B, respectively.  All lattices have $\mpi L\ge 4$, such that finite-volume
effects are sufficiently exponentially suppressed at the level of accuracy
we are working at.

Our data analysis takes into account correlations
between different observables as well as intrinsic autocorrelations of the
Hybrid Monte Carlo (HMC) algorithm resulting from slow modes in the simulation.
As these contributions rapidly grow towards the continuum limit, it is mandatory
to estimate and include a-priori unknown long-tail contributions from the
autocorrelation function of each individual observable. 
Further details of our analysis method can be found in
Appendix~\ref{sec:stat-err-auto}.

\subsection{Extrapolation to the physical point} \label{sec:extrapolation}
To extrapolate our data to the continuum limit and to the physical point, we
take expressions from heavy meson chiral perturbation theory (HM$\chi$PT)
if available, and a linear ansatz otherwise.

In the chiral regime, the mass of the B-meson can be extrapolated to the
physical point using a functional form motivated by
HM$\chi$PT~\cite{Bernardoni:2009sx}. Defining a subtracted mass by removing the
leading non-analytic (in the quark mass) term, viz.  
\begin{equation}
\label{eq:mBsub}
  m^{\rm sub}_{{\rm B}_q,\delta}(y,a) 
  = \mBq(\mpi,a) + c_{q} \frac{3\widehat{g}^2}{16\pi}\left( \frac{\mpi^3}{\fpi^2}-\frac{(\mpi^\xp)^3}{(\fpi^\xp)^2} \right) \;,
\end{equation}
with $c_{q}=1$ in HM$\chi$PT at NLO for $q={\rm d}$ but zero otherwise, the parameterization reads
\begin{equation}
\label{eq:ansatz-mBsub}
  m^{\rm sub}_{{\rm B}_q,\delta}(y,a) 
       = B_{q} + C_{q} \left(y-y^{\rm exp}\right)  + D_{q,\delta} a^2 \,. 
\end{equation}
In \eqref{eq:mBsub} the $B^*\!B\pi$-coupling $\widehat{g} = 0.492(29)$ is taken
from~\cite{Bernardoni:2014kla} and  $y = \mpi^2/(8\pi^2\fpi^2)$ (with the
convention $\fpi^\xp=130.4\,\MeV$ and $\mpi^\xp=134.98\,\MeV$). We use the same
set of measurements for $\fpi$ and $\mpi$ on each CLS ensemble as reported in
foregoing analyses~\cite{Bernardoni:2013xba,Bernardoni:2014fva}. We add the
subscript $\delta$ to distinguish the two available static discretizations
which are combined in the extrapolation to obtain the parameter $B_q\equiv
\mBq$ at the physical point $(y,a)=(y^\xp,0)$. For the $\Bq$-meson mass an
extrapolation quadratically in the lattice spacing is justified as we have full
$\Or(a)$ improvement at work in~\eqref{eq:mBq} at the static order and $\Or(a)$
terms are then suppressed by a factor $1/m_{\rm b}$ once NLO terms in HQET are
taken into account.  For the case of $m_{\rm B}$ that has been explicitly
checked in~\cite{Bernardoni:2013xba} and it is therefore conceivably valid also
here for $m_{\rm B_{\rm s}}$.  

In Fig.~\ref{fig:mBs_Deltasd} we show the continuum and chiral extrapolations 
of $\mBs$ and $\mBs -\mB$ at next-to-leading order in $\minv$.
In this and the following Figures, filled symbols and dashed
curves represent our HYP1 data set, while open symbols and dash-dotted curves
represent our HYP2 data set. For both, equal colours and symbols refer to the
same lattice spacing as indicated by equal values of the bare gauge coupling,
given by $\beta=6/g_0^2$ in the legend, c.f.  Tables~\ref{tab:omega_zb}
and~\ref{tab:meas-parms}. The solid black line is the 
continuum limit given by the part of the fit function
which is independent of the lattice spacing
$a$. Accordingly, the difference between the solid and non-solid lines 
represents the cutoff effect as fitted through a given ansatz.

For the hyperfine splitting, we can perform the chiral and continuum
extrapolation using
\begin{align}\notag
   \dmHF{q,\delta}(y,a) &= [\mBstarq-\mBq]\left[ 1 - 
                           \bar c_{q} \tfrac{3}{2}\widehat{g}^2\left(y\ln{y}-y^{\rm exp}\ln{y^{\rm exp}}\right) + 
                           \bar C_{q} \left(y-y^{\rm exp} \right)  \right] \\
                           &\qquad\qquad\qquad\qquad  + \bar D_{q,\delta} \, a + \bar{\bar {D}}_{q,\delta}\, a^2 \;,
\label{eq:ansatzHF}
\end{align}
with the continuum part coinciding, in the case $q = \rm d$, with the
expression  derived in~\cite{Jenkins:1992hx}. Hence, we can probe 
two ans\"atze for the chiral extrapolation by setting $\bar c_{\rm d}=0,1$.
Since in principle the $\Or(a)$ improvement of the hyperfine
splitting is not implemented, we have included linear cutoff effects,
but also study the scaling behaviour with an $\Or(a^2)$ ansatz by
setting either $\bar D_{q,\delta}$ or $\bar{\bar {D}}_{q,\delta}$ to zero 
in the equation above. In general, our data is not sensitive enough
to clearly separate $\Or(a)$ scaling from $\Or(a^2)$ such that individual 
fits lead to similar results. As additional safety measure we account for
a systematic error by increasing the uncertainty of the favoured $\Or(a)$
extrapolation in order to cover the mean value obtained from the corresponding
$\Or(a^2)$ extrapolation. As will be seen in the following, this only 
occurs in the case of $\mBstars-\mBs$.

Since, to our knowledge, no systematic HM$\chi$PT formulae exist in the
literature for the mass splittings $\mBrq - \mBq$, we again employ a 
simple linear ansatz in $m^2_\pi$: 
\begin{align}\label{eq:ansatzR2}
    \dmradstat{q}(y,a) &= [\mBrq- \mBq]^\stat + C^\stat_{q} (y-y^\xp) + E^\stat_{q,\delta} a^2  \;, \\\label{eq:ansatzR1}
    \dmrad{q}(y,a)     &= [\mBrq- \mBq]       + C'_{q}      (y-y^\xp) + E'_{q,\delta} a +  E''_{q,\delta} a^2 \;.
\end{align} 
For the mass splitting $\dmrad{q}$ at next-to-leading order in HQET, one can in fact define
\begin{align}\label{eq:ansatzR2b}
    \dmradstatminus{q}(y,a) &\equiv \dmrad{q}(y,a) - \dmradstat{q}(y,a)
\end{align}
and take the continuum limit of the leading (static) and next-to-leading
order piece (1/m) separately. While for the former eq.~\eqref{eq:ansatzR2}
can be employed, the $\minv$-correction term has to be extrapolated using 
ansatz~\eqref{eq:ansatzR1} with $E''_{q,\delta}=0$. Both pieces can again
be combined after the physical point is reached, leading to an improved
signal that is dominated by the static extrapolation. No matter what way
is pursued, all extrapolations lead to results that are well compatible 
within errors.

We perform the full analysis using two static discretizations
(HYP1,HYP2) while only one may already suffice. The reason is that the
universality of the continuum limit, a property that can be rigorously
guaranteed for renormalizable theories, implies that both static
discretizations have to lead to equal results in the continuum. At finite
lattice spacing the two discretizations have inherently different lattice
artifacts, best seen among our observables in Fig.~\ref{fig:CL-Ch-extr-HF-split}.
To this end one can either take the two independently in order to show that
universality holds or---as we usually do---combine both in one fit in order to
stabilise the $\chi^2$-minimization. It furthermore serves as a non-trivial
check that the whole procedure has been applied correctly and consistently.

%% file: tables/table_omega_zb.tex
\begin{tabular}{lcccccc}\toprule
     & \multicolumn{2}{c}{$\beta=5.5$} & \multicolumn{2}{c}{$\beta =5.3$} & \multicolumn{2}{c}{$\beta = 5.2$}  \\ 
       \cmidrule(lr){2-3}\cmidrule(lr){4-5}\cmidrule(lr){6-7}
                       & HYP1         & HYP2         & HYP1         & HYP2         & HYP1         & HYP2          \\ \midrule
 $a\mhbare^{\rm stat}$ & $0.969(12)$  & $1.000(12)$  & $1.317(15)$  & $1.350(15)$  & $1.520(17)$  & $1.553(17)$  \\ \cmidrule(lr){2-7}
   $am_{\rm bare}$     & $0.594(14)$  & $0.606(14)$  & $0.993(17)$  & $1.014(17)$  & $1.214(18)$  & $1.239(18)$  \\ 
   $\omegakin/a$       & $0.520(11)$  & $0.525(10)$  & $0.415(8) $  & $0.418(8) $  & $0.377(7) $  & $0.380(7) $  \\ 
   $\omegaspin/a$      & $0.949(37)$  & $1.090(42)$  & $0.73(28) $  & $0.883(33)$  & $0.655(24)$  & $0.812(30)$  \\ 
\bottomrule
\end{tabular}

%% file: tables/table_measparms.tex
%
%

\begin{tabular}{lcccccccccccc} \toprule
     $e$-id & $\beta$ 
         & $\kappa_{\rm d}=\kappa_{\rm sea}$ 
         & $\kappa_{\rm s}$
         & $\tau$
         & $\tau_{\rm cfg}$ 
         & $\tau_{\rm exp}$ 
         & \multicolumn{2}{c}{heavy-light}
         & \multicolumn{2}{c}{heavy-strange}
         \\\cmidrule(lr){8-9}\cmidrule(lr){10-11}
      & & & & & & & HYP1 & HYP2 & HYP1 & HYP2 
         \\ \midrule
      A4 & 5.2 & 0.135900  & 0.1352850 & 2   & 8  & 106 & \multicolumn{2}{l}{\quad 1012   [1,1]} & \multicolumn{2}{l}{\;\quad 1000 [1,1] }       \\
      A5c&     & 0.135940  & 0.1352570 & 2   & 4  & 39  & \multicolumn{2}{l}{\quad 501    [1,1]} & \multicolumn{2}{l}{\;\quad 500  [1,1]  }       \\
      A5d&     &           &           &     &    &     & \multicolumn{2}{l}{\quad 500    [1,1]} & --           &  --                           \\
      B6 &     & 0.135970  & 0.1352390 & 2   & 2  & 39  & \multicolumn{2}{l}{\quad 636    [1,1]} & \multicolumn{2}{l}{\;\quad 636  [1,1]  }       \\[0.2em]
      E5 & 5.3 & 0.136250  & 0.1357770 & 4   &16  & 151 & \multicolumn{2}{l}{\quad 1000   [1,1]} & \multicolumn{2}{l}{\;\quad 1000 [1,1] }       \\
      F6 &     & 0.136350  & 0.1357410 & 2   & 8  & 151 & \multicolumn{2}{l}{\quad 500    [1,1]} & \multicolumn{2}{l}{\;\quad 600  [1,1]  }       \\
      F7 &     & 0.136380  & 0.1357300 & 2   & 8  & 151 & \multicolumn{2}{l}{\quad 602  [100,1]} & \multicolumn{2}{l}{\;\quad 200  [1,2]  }       \\
      G8 &     & 0.136417  & 0.1357050 & 2   & 2  & 56  & \multicolumn{2}{l}{\quad 410    [1,1]} & --           &  --                           \\[0.2em]
      N5 & 5.5 & 0.136600  & 0.1362620 & 0.5 & 8  & 488 & \multicolumn{2}{l}{\quad 477  [300,1]} & --           &  --                           \\
      N6 &     & 0.136670  & 0.1362500 & 2   & 4  & 108 & \multicolumn{2}{l}{\quad 950    [2,2]} & \multicolumn{2}{l}{\;\quad 707 [100,2]}       \\
      O7 &     & 0.136710  & 0.1362430 & 2   & 4  & 108 & \multicolumn{2}{l}{\quad 980    [1,1]} & --           & \multicolumn{1}{l}{490 [1,2]} \\
    \bottomrule                               
\end{tabular}

%% file: s3.tex
\section{Results}

Using the HQET parameters of Table~\ref{tab:omega_zb} at $\zb=13.25$, we obtain
for the B-meson mass $\mB = 5.285(62)$\,GeV by employing
eqs.~\eqref{eq:mBsub}~and~\eqref{eq:ansatz-mBsub} with $c_q=1$ as extrapolation
ansatz. Note, however, that the experimental value $\mB=5.2795$~GeV has been
used as input in~\cite{Bernardoni:2013xba} to fix the b-quark mass. Therefore
$\mB$ is not a prediction of the theory in our setup, and the number above
should be regarded as a consistency check of the approach.

\subsection{Ground states}

\begin{figure}[t]
        \centering
        \includegraphics[height=0.355\textwidth]{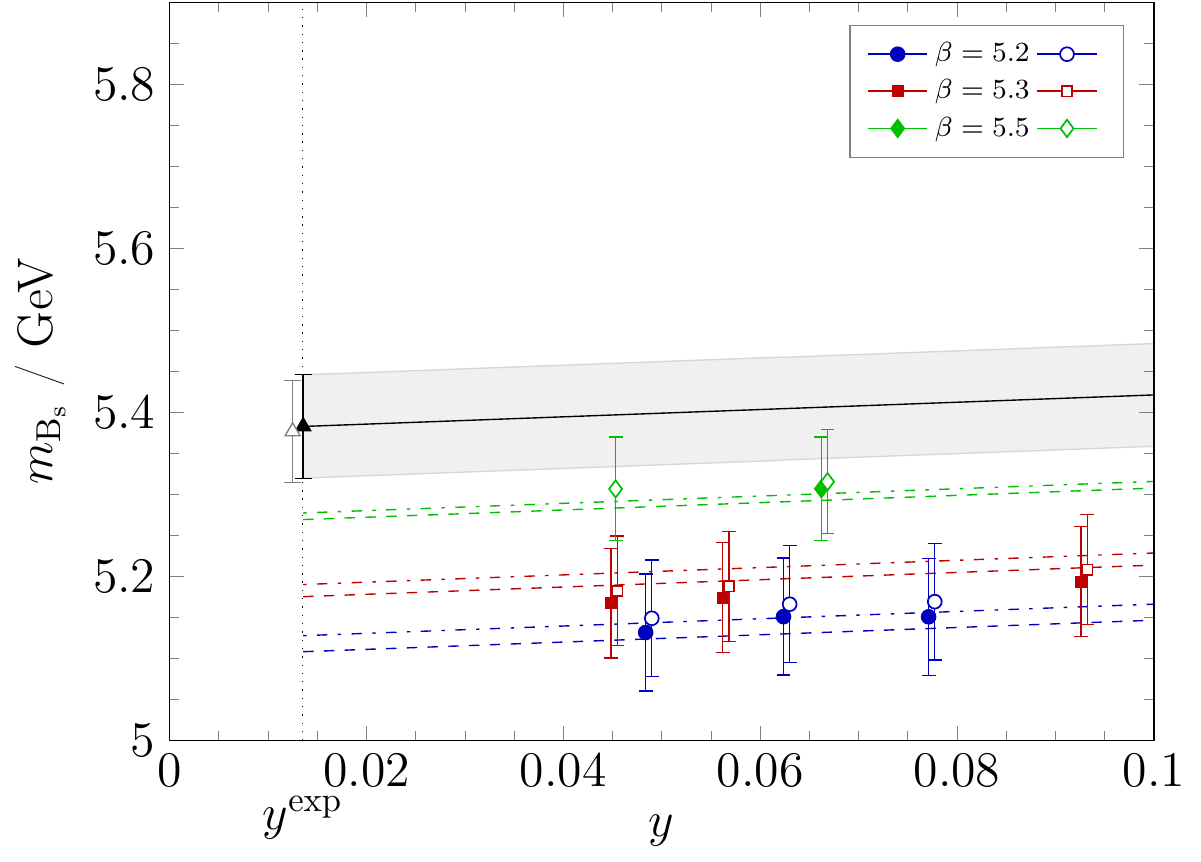}\hfill
        \includegraphics[height=0.355\textwidth]{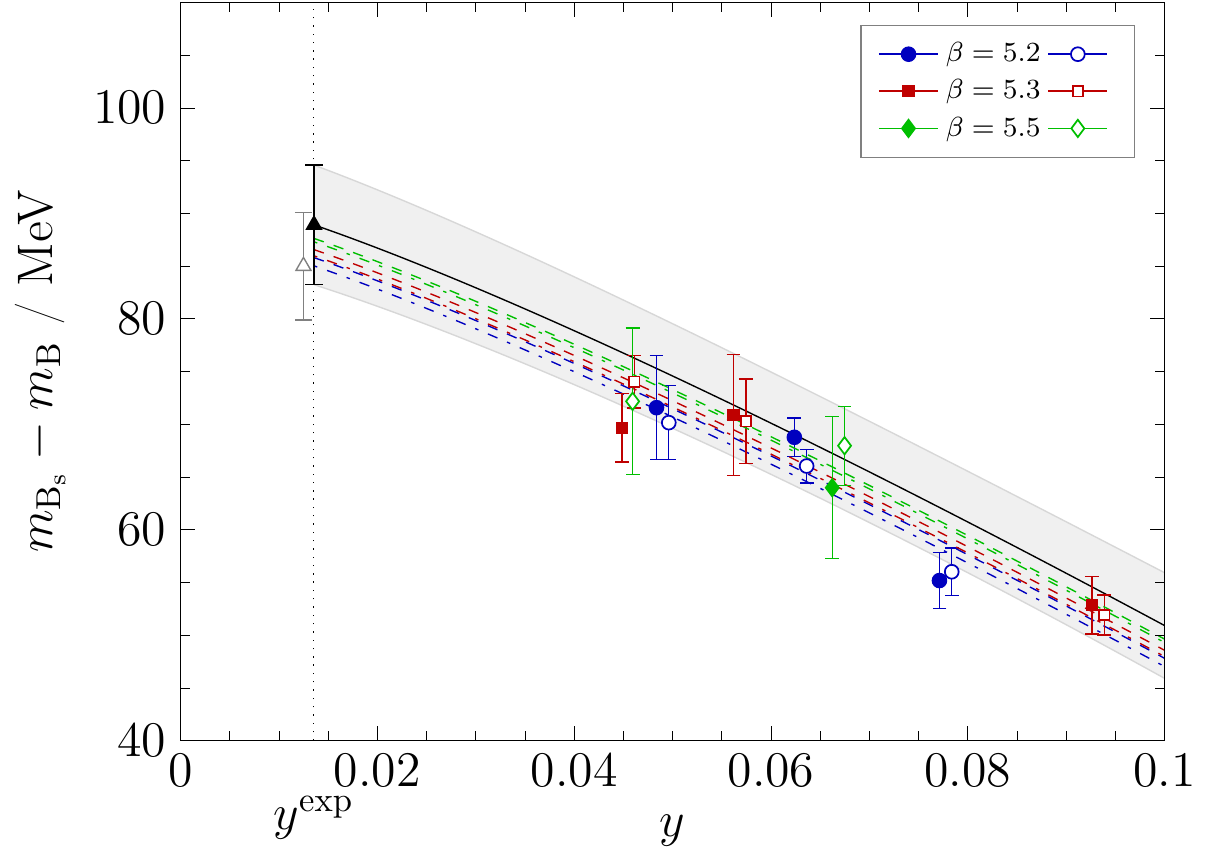}
        \vskip-1em
        \caption{Chiral and continuum extrapolation of $\mBs$
                 (\textit{left}) and $\mBs-\mB$ ({\it right}) according
                 to~\eqref{eq:ansatz-mBsub}.
                 The open triangle represents the corresponding result
                 for extrapolating the static order data. 
                 Here and in the following Figures, filled symbols 
                 and dashed curves represent our HYP1 data set, while open symbols 
                 and dash-dotted curves represent our HYP2 data set. The solid 
                 black line is the continuum limit for the given fit ansatz.
        }
        \label{fig:mBs_Deltasd}
\end{figure}

As first quantities beyond $\mB$, we compute the $\Bs$-meson mass
and the mass difference $\mBs -\mB$. Their continuum and chiral extrapolations
at next-to-leading order in $\minv$ are shown in Fig.~\ref{fig:mBs_Deltasd}
and compared to the extrapolated value of the static data at the physical
point. The raw data for the mass difference $\mBs - \mB$ and the hyperfine 
splittings is collected in Tables~\ref{tab:sd-split-raw} and~\ref{tab:hf-split-raw}.

From the extrapolation of the form of eqs.~\eqref{eq:mBsub}~and~\eqref{eq:ansatz-mBsub} 
with $c_q=0$, we obtain the result:
\begin{align}\label{eq:Bs-mass}
   \mBs             &=  5383(63) \,\MeV  \;, &
   \mBs-\mBs^\stat  &=  58(12) \,\MeV  \;,
\end{align}
where the error includes statistic and systematic uncertainties (scale setting,
HQET parameters), combined in quadrature as explained in 
Appendix~\ref{sec:stat-err-auto}.
Although our result for $\mBs$ comes with a much larger error compared to the PDG value, 
$m_{\rm B^0_s}=5366.77\pm0.24\,\MeV$, the difference between the
mean values is only one fourth of our error. 

In the combination $\mBs-\mB$, the error is reduced due to correlations among the 
heavy-light and heavy-strange measurements, although the latter have been performed
on a subset of the available ensembles only, as it can be inferred
from Table~\ref{tab:meas-parms}. 

Our results for the $\Bs$-B mass splitting at $\Or(1/m)$ and
in the static approximation are
\begin{subequations}\label{eq:BsBmass-split}
\begin{align}\label{eq:BsBmass-split-HQET}
    \mBs - \mB           &= 88.9(5.7)(2.3)_{\chi} \,\MeV    \;, \\
             \label{eq:BsBmass-split-stat}
   [\mBs - \mB ]^{\stat} &= 85.0(5.1)(2.2)_{\chi} \,\MeV \;,
\end{align}
\end{subequations}
both in good agreement with the PDG value of
\cite{PDG}
$\mBs - \mB = 87.35(23) \,\MeV$. Here, the quoted mean and statistical error
results from an HM$\chi$PT extrapolation ansatz that is obtained by
appropriately combining eq.~\eqref{eq:ansatz-mBsub} for the light and strange
quark sector. As systematic error due to the chiral extrapolation ansatz we
quote its difference to the standard $\mpi^2$-extrapolation.  

\begin{figure}[t]
        \centering
        \includegraphics[height=0.355\textwidth]{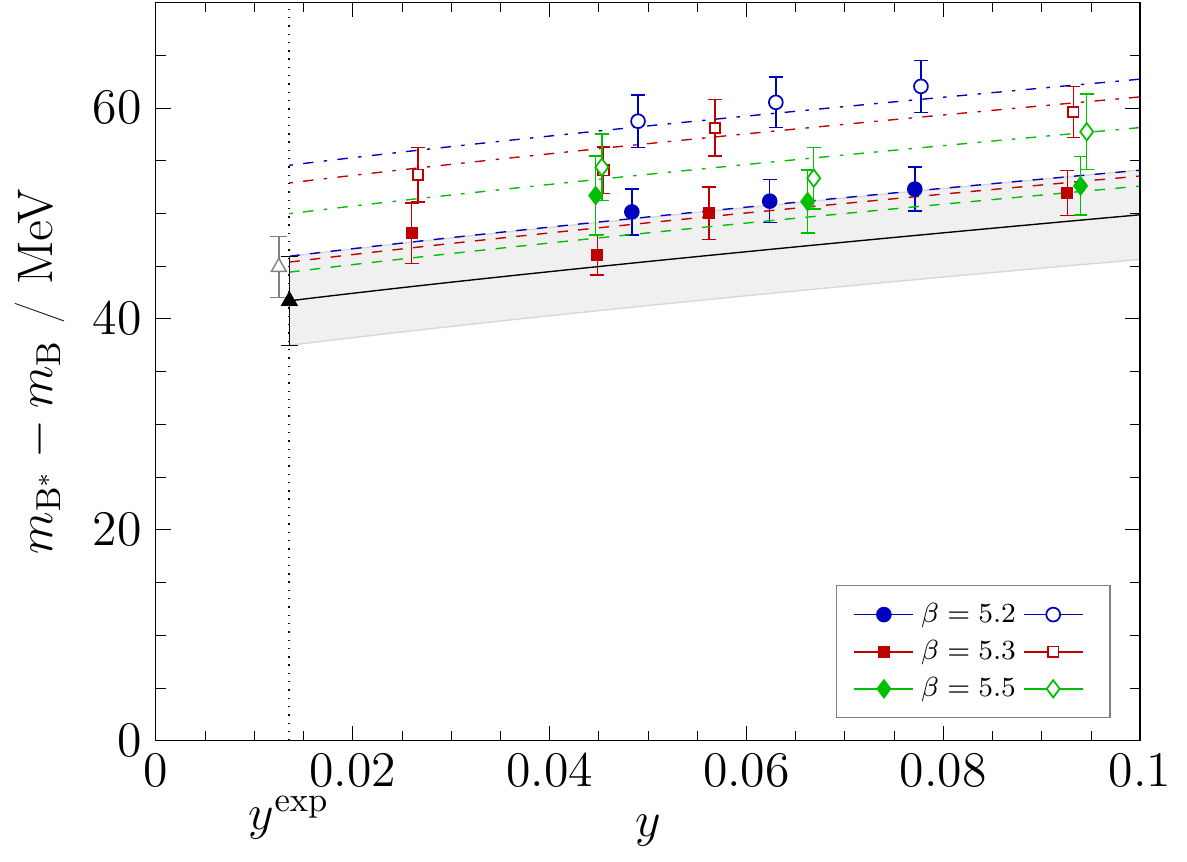}
        \includegraphics[height=0.355\textwidth]{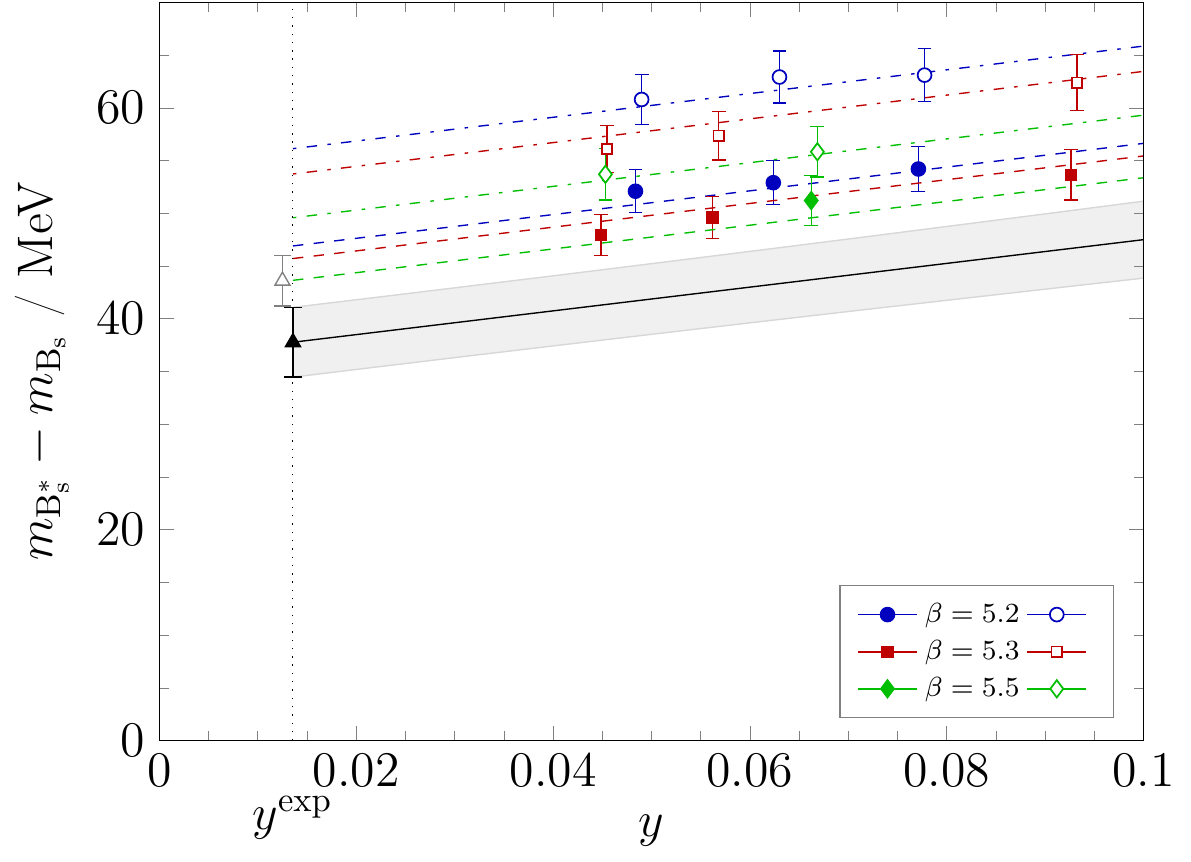}
        \vskip-1em
        \caption{Chiral and continuum extrapolation of the hyperfine spin
                 splitting of the B-meson ({\it left}) and  
                 of the $\Bs$-meson ({\it right}). For $\mBstar-\mB$
                 we show the extrapolation from ansatz~\eqref{eq:ansatzHF}
                 with $\bar c_{\rm d}=1$ and leading $a$-effects, while for
                 $\mBstars-\mBs$ we have to set $\bar c_{\rm s}=0$. In both
                 panels the open triangle at the physical point reflects 
                 the corresponding fit result using an $a^2$-scaling ansatz. 
                 Details are given in Table~\ref{tab:sd-split-raw}.
        }
        \label{fig:CL-Ch-extr-HF-split}
\end{figure}
\begin{table}[t]
  \centering\small
  \input{tables/table_sdsplit.tex}
  \caption{Mass splitting between $\Bs$- and $\Bd$-meson at static and next-to-leading order HQET.}
  \label{tab:sd-split-raw}
\end{table}
\begin{table}[t]
  \centering\small
  \input{tables/table_hfsplit.tex}
  \caption{Hyperfine splittings in the light- and strange-quark sector.}
  \label{tab:hf-split-raw}
\end{table}

For the hyperfine spin splittings, we obtain 
\begin{subequations}\label{eq:hf-split}
\begin{align}\label{eq:hf-split-l}
    \mBstar  - \mB           &= 41.7(4.2)(3.2)_{a}(0.3)_{\chi}  \,\MeV \;,\\ 
             \label{eq:hf-split-s}
    \mBstars - \mBs          &= 37.8(3.3)(5.8)_{a}              \,\MeV \;,
\end{align}
\end{subequations}
where we quote the mean value obtained with $\bar c_{\rm d}=1$ for the
B-meson case and add the difference w.r.t. the result obtained from the 
ansatz with $\bar c_{\rm d}=0$ as a systematic error estimate for the
chiral extrapolation.
As mentioned earlier, we also account here for a systematic error between linear and 
quadratic continuum extrapolations. The $\Or(a)$ extrapolations 
are shown for both the B and the $\Bs$ system in Fig.~\ref{fig:CL-Ch-extr-HF-split}.
The filled/empty symbols at the physical
point are the results using either an $\Or(a)$, or an $\Or(a^2)$ term in
the continuum extrapolation. 

Our result for the hyperfine splitting for the B-system is in good agreement with the
experimental value $\mBstar - \mB=45.78(35) \,\MeV$ \cite{PDG},
whereas the $\Bs$ hyperfine splitting differs
noticeably from the experimental value $\mBstars - \mBs = 48.7{+2.3\atop-2.1}\,\MeV$.
Our results are smaller than the experimental value and than our result for the 
hyperfine splitting of the B (the opposite of the situation for the experimental values).
Since the hyperfine splitting came out far too small in the quenched approximation 
\cite{Blossier:2010vz},
this is suggestive of a residual quenching effect from the quenching of the 
strange quark in our $\Nf=2$ simulations. Moreover, Fig.~\ref{fig:CL-Ch-extr-HF-split}
indicates larger cutoff effects in the case of the hyperfine splitting for the $\Bs$.

\subsection{Excited states}\label{sec:excited}

\begin{table}[p]
  \centering\small
  \input{tables/table_radsplit_hl.tex}

  \caption{Mass gaps between the excited state $\Brd$- and the $\Bd$-meson at static and next-to-leading order HQET.}
  \label{tab:radsplit-hl}
\end{table}

\begin{table}[p]
  \centering\small
  \input{tables/table_radsplit_hs.tex}

  \caption{Mass gaps between the excited state $\Brs$- and the $\Bs$-meson at static and next-to-leading order HQET.}
  \label{tab:radsplit-hs}
\end{table}
\begin{table}[p]
  \centering\small
  \input{tables/table_radsplit_1ovm.tex}

  \caption{Subleading HQET contributions to the mass gaps between excited and ground states.}
  \label{tab:radsplit-1ovm}
\end{table}

\begin{figure}[p]
        \centering
        \includegraphics[height=0.35\textwidth]{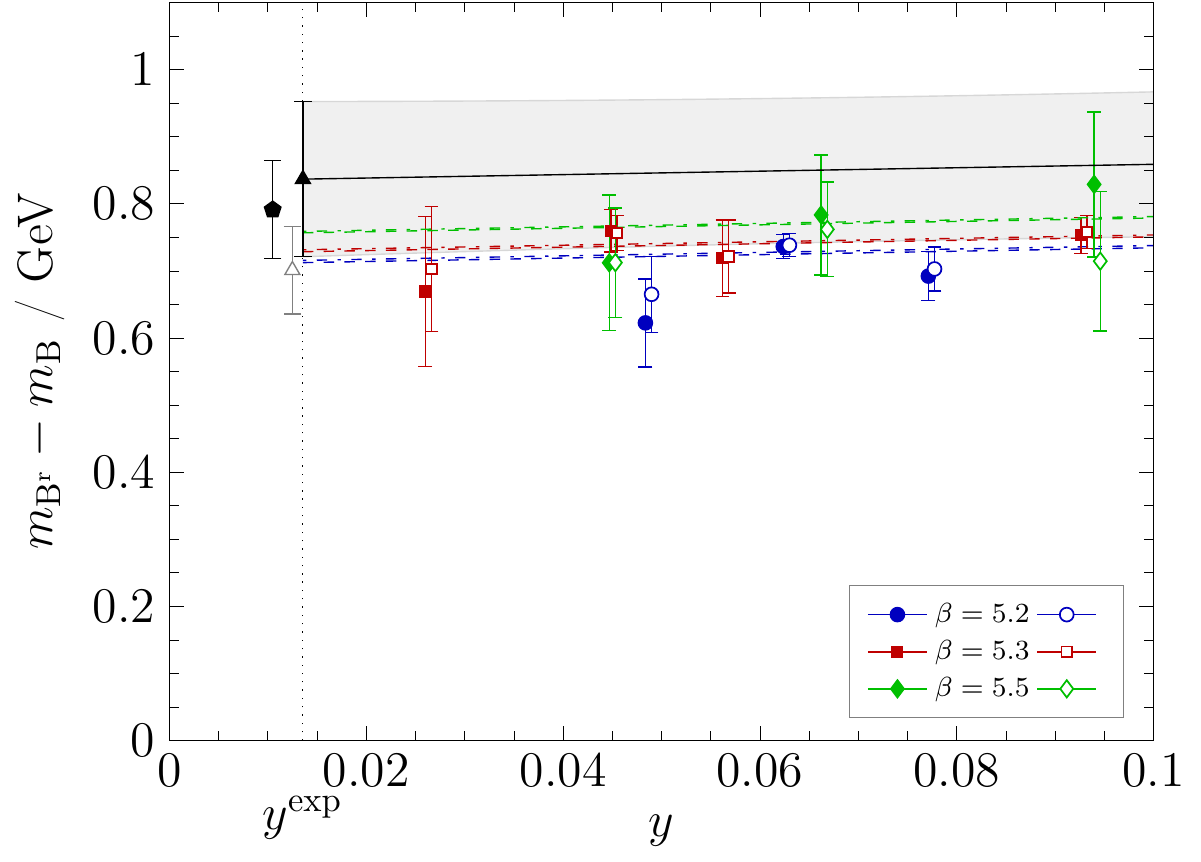}\hfill
        \includegraphics[height=0.35\textwidth]{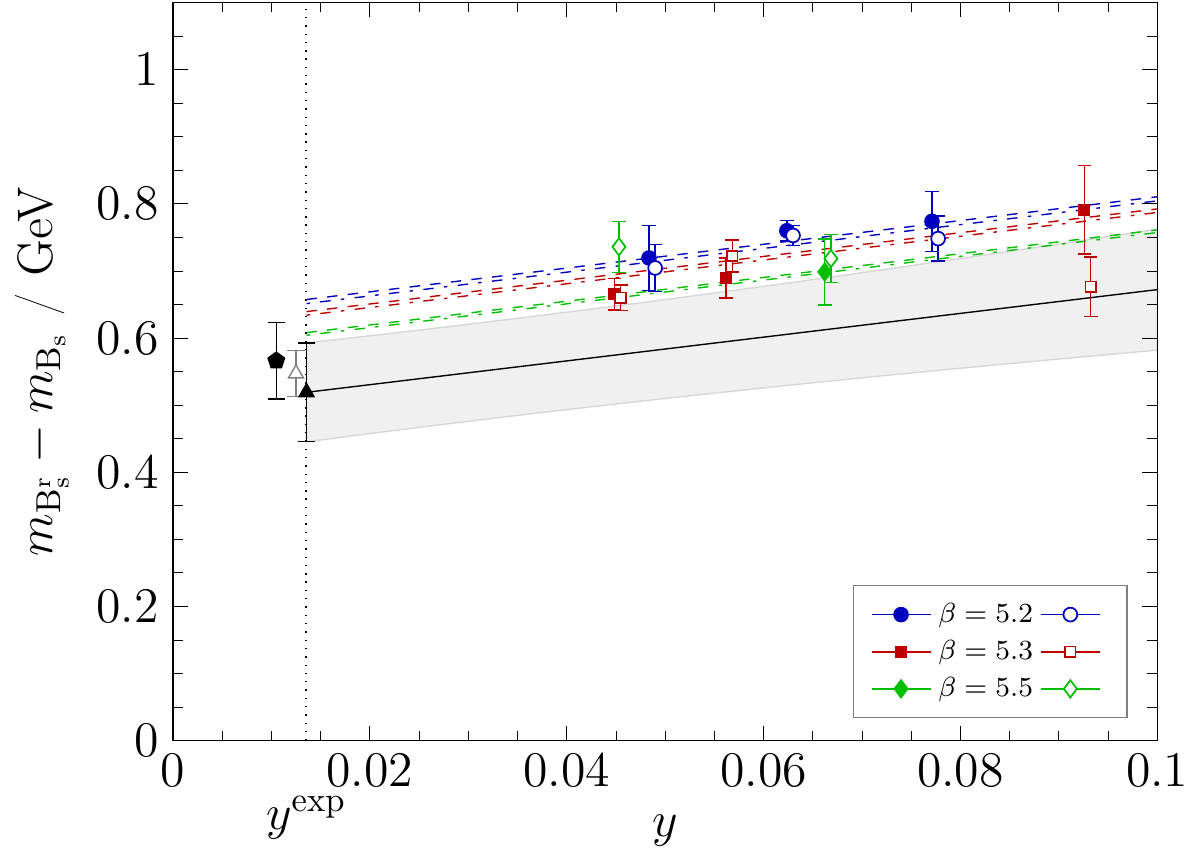}
        \vskip-1em
        \caption{Chiral and continuum extrapolation of $\mBr-\mB$
                 (\textit{left}) and $\mBrs-\mBs$ ({\it right}).  Both
                 represent an extrapolation with ansatz~\eqref{eq:ansatzR1} and
                 $E''_{q,\delta}=0$ for HQET to order $\minv$.  The open
                 triangle represents the corresponding result for extrapolating
                 the static order data. For details see
                 Tables~\ref{tab:radsplit-hl} and~\ref{tab:radsplit-hs}.  We
                 also add the continuum point (filled black pentagon)
                 corresponding to the combination given in
                 eq.~\eqref{eq:B-radsplit}.
                }
        \label{fig:CLradsplit}
\end{figure}

\begin{figure}[t]
  \centering
  \includegraphics[height=0.46\textwidth]{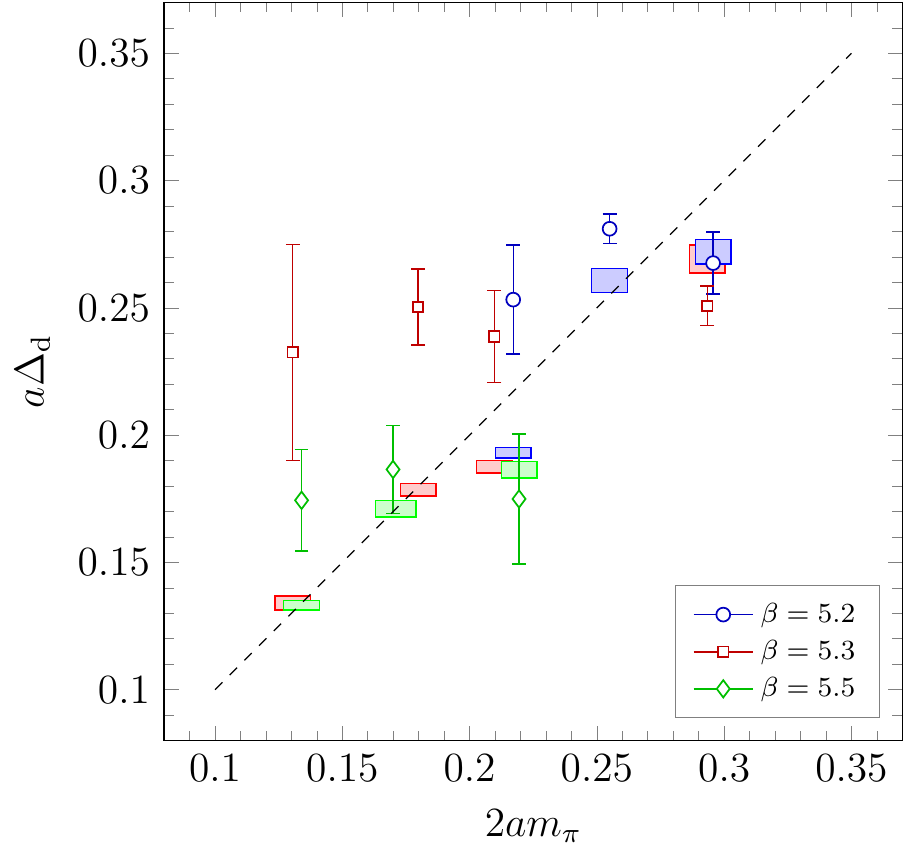}\hfill
  \includegraphics[height=0.46\textwidth]{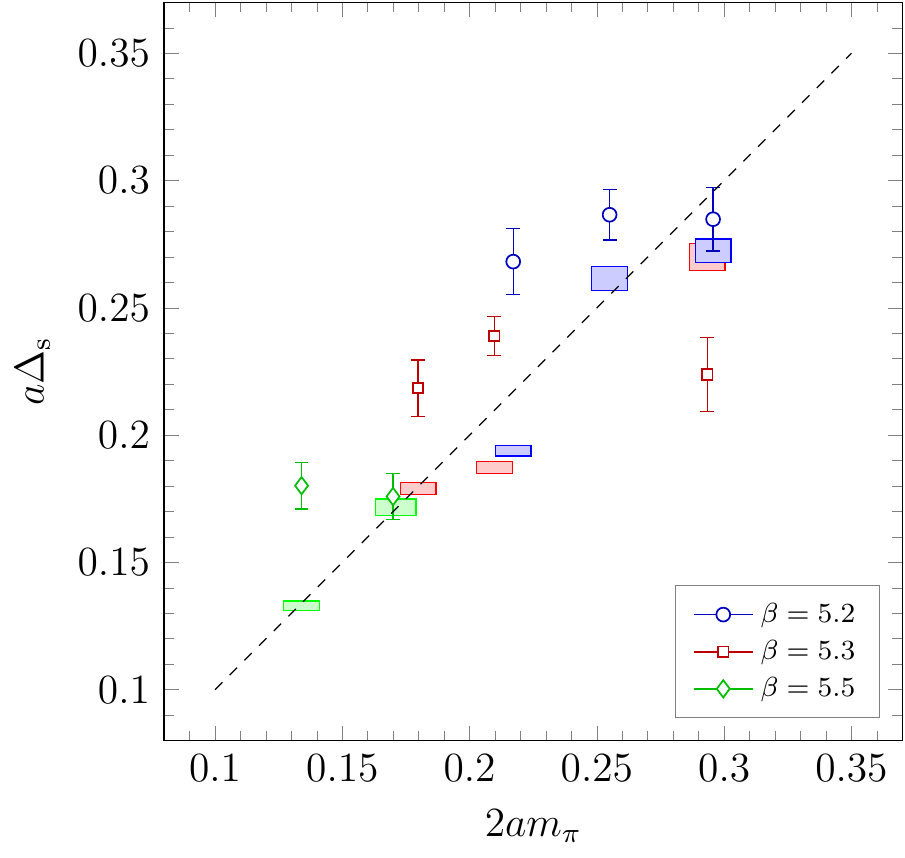}
  \vskip-1em
  \caption{The measured gap $\mBrq - \mBq$ vs. $2m_\pi$ for the excited states
           $\Brq$ with $q={\rm d}$ (\textit{left}) and $q={\rm s}$
           (\textit{right}) on each individual CLS ensemble and with HYP2
           discretization of the static quark.  The boxes are an estimate of
           the expected gap for a two-hadron state $\Bstarq(-p)+\pi(p)$ with
           one unit of lattice momentum $p=|{\bf p}|=2\pi/L$.  The lower edge
           of the boxes corresponds to $\Delta^{\rm HF}_q\!m + \sqrt{m_\pi^2 +
           (2\pi/L)^2}$ and the upper edge includes an additional contribution
           ${\bf p}^{2}/(2\mBstarq)$ as a naive estimate for the kinetic energy of the
           $\Bstarq$.  The dotted line is $\Delta=2m_\pi$ corresponding to a
           three-hadron state $\Bq + \pi + \pi$.
  }
  \label{fig:gaps_vs_2mpi}
\end{figure}
\begin{figure}[t]
  \centering
  \includegraphics[height=0.46\textwidth]{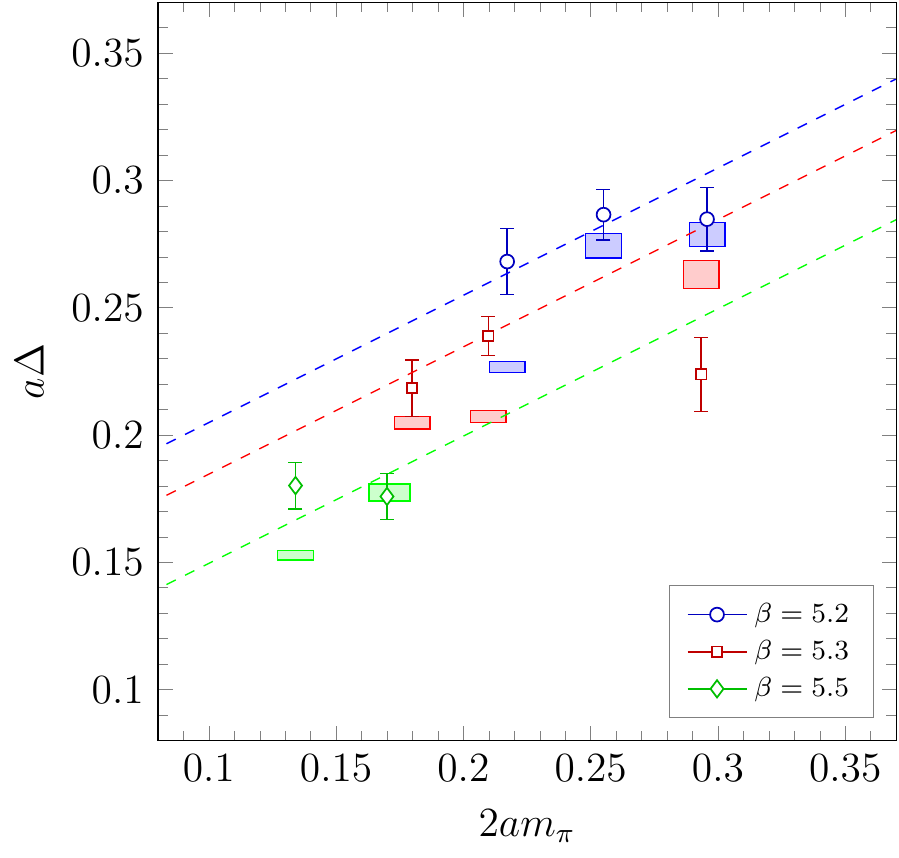}
  \vskip-1em
  \caption{The same data points as in the right panel of Fig.~\ref{fig:gaps_vs_2mpi}
           for the measured $\Brs$ mass gaps, together with estimates of the energy 
           gap for a two-hadron state $\Bstar(-p) + \mathrm{K}(p)$ (boxes) and a 
           three-hadron state $\Bd+\mathrm{K}+\pi$ (dashed lines). 
  }
  \label{fig:gapK_vs_2mpi}
\end{figure}

For the mass splittings between the ground state $\Bq$ and the first excited state, 
denoted here by $\Br$ and $\Brs$, we obtain
\begin{subequations}\label{eq:B-radsplit}
\begin{align}\label{eq:B-radsplit-HQET}
    \mBr - \mB           &= 791(73)    \,\MeV \;, &
   [\mBr - \mB ]^{\stat} &= 701(65)    \,\MeV \;, \\
             \label{eq:Bs-radsplit-HQET}
    \mBrs -\mBs          &= 566(57)    \,\MeV \;, &
   [\mBrs -\mBs]^{\stat} &= 547(34)    \,\MeV \;.
\end{align}
\end{subequations}
after adding the individually extrapolated results for $[\mBrq - \mBq ]^\stat$ 
and $[\mBrq - \mBq ]^{\rm 1/m}$.
In Fig.~\ref{fig:CLradsplit} these values are shown (as pentagons, slightly shifted 
at the physical point) for comparison with an extrapolation according to 
eq.~\eqref{eq:ansatzR1}.
The raw data is collected in Tables~\ref{tab:radsplit-hl},~\ref{tab:radsplit-hs}, 
and~\ref{tab:radsplit-1ovm}.

We conclude this Section by remarking that for the excited states the 
interpretation of our results in terms of mass differences of physical 
one-meson states, e.g. ``radial excitations'', is not straight forward.
Although our values for these mass gaps are larger than what a multi-hadron 
state made of, e.g., a B$^{(*)}_{(\rm s)}$-meson and a small number ($\leq 2$) of 
physical pions would produce, we cannot unambiguously conclude that our 
$\Brq$ states actually correspond to radial excitations of the ground-state 
$\Bq$-mesons. 
In a rigorous approach, states above multi-hadron thresholds need to be 
treated as resonances in order to obtain precise values for their masses
and widths, and to associate them with the states observable in experiments.

Since our lattice study is unquenched, the excited states can also be 
multi-particle states involving additional pions, beside the desired 
one-particle state. 
While it has been argued that the overlap of single-hadron interpolating 
operators to multi-hadron states is small~\cite{Michael:2005kw,Bar:2012ce}, 
the two-hadron states may have a weaker volume suppression \cite{Bar:2015zwa}.
Moreover, a chiral extrapolation linear in $m^2_\pi$ might not 
be adequate for a multi-hadron state, and a different extrapolation
ansatz for these states, e.g. linear in $m_\pi$, might also yield somewhat 
smaller mass gaps.

In continuum (and infinite-volume) physics, any excited $\Bq$-meson state, 
$\Bxq$, which strongly decays into an $\ell$-wave $\Bq +\pi$ state, i.e. a resonance 
in the $\ell$-wave $\Bq + \pi$ scattering channel, implies that the corresponding 
two-hadron state $\Bxq+\pi$ with relative angular momentum $\ell$ has the correct 
quantum numbers to couple to our interpolating operators used for the $\Bq$.
For the $q={\rm d}$ sector, the set of possible two-hadron excited states includes
\begin{itemize}
\item ${\rm B}^\ast + \pi$ in P-wave which would have a non-interacting two-hadron 
      energy gap of $\Delta \gtrsim 181\,\MeV$,
\item ${\rm B}^{\prime\ast} + \pi$ in P-wave, where ${\rm B}^{\prime\ast}$ 
      is a radial excitation ($J^P=1^-$), 
\item ${\rm B}_0^\ast + \pi$ in S-wave, where ${\rm B}_0^\ast$ is
      an orbital excitation ($J^P=0^+$),
\item ${\rm B}_2^\ast + \pi$ in D-wave, where ${\rm B}_2^\ast$ is 
      an orbital excitation ($J^P=2^+$), like the observed ${\rm B}_2^\ast(5747)^0$ state
      which would lead to a non-interacting two-hadron energy gap of $\Delta \gtrsim 598\,\MeV$.
\end{itemize}
On a finite lattice, the energies of states with non-zero relative 
angular momentum are lifted due to the minimal momentum of $|{\bf p}|=2\pi/L$. 
Since the s-quark is quenched in our study, we cannot have 
excited states in the B sector with an s-quark from the sea, 
like $\Bstars + \textrm{K}$. 
On the other hand, in the $\Bs$ sector there are two flavour combinations 
for each of the corresponding excited state of the B sector, e.g.
$\Bstars+\pi$ and $\Bstar +\textrm{K}$ (corresponding to $\Bstar+\pi$).

In Figs.~\ref{fig:gaps_vs_2mpi} and~\ref{fig:gapK_vs_2mpi} we show
our data points on the various ensembles together with an 
estimate for the gaps of some two-hadron states. The comparison 
indicates that some of the two-hadron states can be close in energy 
to the excited states we measured.
Depending on the pion mass on a given ensemble, 
the energy gaps determined according to eqs.~\eqref{eq:radial} may in 
fact be the energy splitting to the lowest lying $\Bx + \pi$ state, 
if this is lighter than the radial excitations $\Bprimeq$. 

Below three- or more-hadron thresholds, the infinite volume scattering matrix 
(up to corrections which vanish exponentially in the spatial lattice extent) 
may be inferred from the finite volume energy 
spectrum~\cite{Luscher:1985dn,Luscher:1986pf,Luscher:1990ux,Luscher:1991cf,Hansen:2012tf}. 
In practice this requires the construction of correlation matrices which
contain two-hadron as well as single hadron operators. To date, this procedure
has been carried out successfully in some simple systems such as $\pi\!-\!\pi$
and $\pi-$K scattering, and the energy dependence of the scattering phase shift
has been determined with a sufficient resolution to clearly discern 
resonant behaviour (see e.g.~\cite{Yamazaki:2015nka} for a recent review). 
Without performing such a dedicated study, we are not able to conclusively 
determine the nature of the excited state we observe.
\begin{table}[t]
  \centering\small
  \input{tables/res_errors.tex}

  \caption{Distributions of relative squared errors,
           $\sigma_{i}^2/\sum_{j}\sigma_{j}^2$ among different sources $i$
           for the mean values given
           in~\eqref{eq:Bs-mass}-\eqref{eq:B-radsplit}. In the second row we
           repeat them in \MeV~for the reader's convenience, including only the
           statistical error of the mean. Remaining relative errors enter
           through the scale setting procedure only.}
  \label{tab:errdist}
\end{table}

%% file: tables/table_sdsplit.tex
%
%
\begin{tabular}{clllll} \toprule
        &     & \multicolumn{2}{c}{$\Delta_{\rm s-d}m$\;[\MeV]}& 
                \multicolumn{2}{c}{$\Delta^{\rm stat}_{\rm s-d}m$\;[\MeV]}   \\ 
       \cmidrule(r){3-4}\cmidrule(r){5-6}
 $e$-id& $y$        & HYP1      & HYP2      & HYP1      & HYP2       \\ \midrule
    A4 & 0.0771(14) & 55.2(2.6) & 56.0(2.3) & 51.9(2.1) & 53.1(1.8)  \\
    A5 & 0.0624(13) & 68.8(1.8) & 66.0(1.6) & 62.6(1.8) & 63.4(1.7)  \\
    B6 & 0.0484(9)  & 71.6(4.9) & 70.2(3.5) & 66.5(3.9) & 67.3(3.1)  \\[0.15em]
    E5 & 0.0926(15) & 52.8(2.7) & 51.9(1.9) & 48.8(2.0) & 49.2(1.7)  \\
    F6 & 0.0562(9)  & 70.9(5.7) & 70.3(4.0) & 66.5(5.2) & 64.7(3.9)  \\
    F7 & 0.0449(8)  & 69.7(3.2) & 74.0(2.5) & 67.1(3.0) & 69.4(2.4)  \\[0.15em]
    N6 & 0.0662(10) & 64.0(6.7) & 68.0(3.7) & 62.8(4.8) & 63.3(3.2)   \\
    O7 & 0.0447(7)  & --        & 72.2(6.9) &    --     & 70.2(5.3)  \\\midrule
    LO-$a^{2}$ & $y^{\rm exp},a=0$ & \multicolumn{2}{c}{91.2(5.8)}
                                   & \multicolumn{2}{c}{87.2(5.2)}  \\
    NLO-$a^{2}$& $y^{\rm exp},a=0$ & \multicolumn{2}{c}{88.9(5.7)}
                                   & \multicolumn{2}{c}{85.0(5.1)}  \\
    \bottomrule
\end{tabular}

%% file: tables/table_hfsplit.tex
%
%
\begin{tabular}{clllll} \toprule
        &     & \multicolumn{2}{c}{$\Delta^{\rm HF}_{\rm d}\!m$\;[\MeV]}& \multicolumn{2}{c}{$\Delta^{\rm HF}_{\rm s}\!m$\;[\MeV]}   \\ 
       \cmidrule(r){3-4}\cmidrule(r){5-6}
 $e$-id& $y$        & HYP1      & HYP2      & HYP1      & HYP2       \\ \midrule
    A4 & 0.0771(14) & 52.3(2.1) & 62.0(2.4) & 54.2(2.2) & 63.1(2.5)  \\
    A5 & 0.0624(13) & 51.1(2.0) & 60.5(2.4) & 52.9(2.1) & 62.9(2.5)  \\
    B6 & 0.0484(9)  & 50.1(2.2) & 58.7(2.5) & 52.1(2.0) & 60.8(2.4)  \\[0.15em]
    E5 & 0.0926(15) & 51.9(2.1) & 59.6(2.4) & 53.6(2.4) & 62.4(2.7)  \\
    F6 & 0.0562(9)  & 50.0(2.5) & 58.1(2.7) & 49.6(2.0) & 57.4(2.3)  \\
    F7 & 0.0449(8)  & 46.1(1.9) & 54.1(2.2) & 47.9(2.0) & 56.1(2.2)  \\
    G8 & 0.0260(5)  & 48.1(2.9) & 53.7(2.6) &   --      &  --        \\[0.15em]
    N5 & 0.0940(19) & 52.6(2.8) & 57.7(3.6) &   --      &  --        \\
    N6 & 0.0662(10) & 51.1(3.0) & 53.3(2.9) & 51.2(2.4) & 55.8(2.4)  \\
    O7 & 0.0447(7)  & 51.7(3.7) & 54.4(3.1) &    --     & 53.7(2.4)  \\\midrule
    LO-$a^{2}$ & $y^{\rm exp},a=0$ & \multicolumn{2}{c}{45.2(2.9)}
                                   & \multicolumn{2}{c}{43.6(2.4)}  \\
    LO-$a^{1}$ & $y^{\rm exp},a=0$ & \multicolumn{2}{c}{42.0(4.2)}
                                   & \multicolumn{2}{c}{37.8(3.3)}  \\
   NLO-$a^{2}$ & $y^{\rm exp},a=0$ & \multicolumn{2}{c}{44.9(2.9)}  \\
   NLO-$a^{1}$ & $y^{\rm exp},a=0$ & \multicolumn{2}{c}{41.7(4.2)}  \\
    \bottomrule
\end{tabular}

%% file: tables/table_radsplit_hl.tex
%
%
\begin{tabular}{clllll} \toprule
           &     
           & \multicolumn{2}{c}{$\dmrad{\rm d}$\;[\GeV]}
           & \multicolumn{2}{c}{$\dmradstat{\rm d}$\;[\GeV]}   \\ 
          \cmidrule(r){3-4}\cmidrule(r){5-6}
    $e$-id& $y$        & HYP1      & HYP2      & HYP1      & HYP2      \\ \midrule
       A4 & 0.0771(14) & 0.692(37) & 0.703(33) & 0.619(27) & 0.625(25) \\
       A5 & 0.0624(13) & 0.736(18) & 0.738(17) & 0.655(16) & 0.664(15) \\
       B6 & 0.0484(9)  & 0.622(66) & 0.665(57) & 0.552(56) & 0.592(52) \\[0.15em]
       E5 & 0.0926(15) & 0.753(27) & 0.758(25) & 0.676(24) & 0.682(23) \\
       F6 & 0.0562(9)  & 0.719(57) & 0.721(54) & 0.658(52) & 0.654(47) \\
       F7 & 0.0449(8)  & 0.760(32) & 0.756(26) & 0.678(32) & 0.676(30) \\
       G8 & 0.0260(5)  & 0.67(11)  & 0.70(9)   & 0.63(11)  & 0.65(9)   \\[0.15em]
       N5 & 0.0940(19) & 0.83(11)  & 0.71(14)  & 0.697(35) & 0.709(31) \\
       N6 & 0.0662(10) & 0.78(9)   & 0.76(8)   & 0.634(72) & 0.668(58) \\
       O7 & 0.0447(7)  & 0.71(12)  & 0.71(8)   & 0.64(10)  & 0.63(8)   \\\midrule
 LO-$a^2$ & $y^{\rm exp},a=0$ & \multicolumn{2}{c}{0.787(71)}
                              & \multicolumn{2}{c}{0.701(65)}  \\
 LO-$a^1$ & $y^{\rm exp},a=0$ & \multicolumn{2}{c}{ 0.84(12)}
                              &                                \\
    \bottomrule
\end{tabular}

%% file: tables/table_radsplit_hs.tex
%
%
\begin{tabular}{clllll} \toprule
           &     
           & \multicolumn{2}{c}{$\dmrad{\rm s}$\;[\GeV]}
           & \multicolumn{2}{c}{$\dmradstat{\rm s}$\;[\GeV]}   \\ 
          \cmidrule(r){3-4}\cmidrule(r){5-6}
    $e$-id& $y$        & HYP1      & HYP2      & HYP1      & HYP2      \\ \midrule
       A4 & 0.0771(14) & 0.774(45) & 0.748(34) & 0.640(23) & 0.650(22) \\
       A5 & 0.0624(13) & 0.760(15) & 0.753(15) & 0.670(11) & 0.677(11) \\
       B6 & 0.0484(9)  & 0.719(49) & 0.704(35) & 0.641(22) & 0.641(21) \\[0.15em]
       E5 & 0.0926(15) & 0.791(66) & 0.676(44) & 0.609(29) & 0.624(25) \\
       F6 & 0.0562(9)  & 0.689(30) & 0.722(24) & 0.643(18) & 0.642(17) \\
       F7 & 0.0449(8)  & 0.665(24) & 0.660(19) & 0.593(15) & 0.601(14) \\[0.15em]
       N6 & 0.0662(10) & 0.698(49) & 0.718(36) & 0.621(38) & 0.643(32) \\
       O7 & 0.0447(7)  & --        & 0.736(38) & --        & 0.669(29) \\\midrule
 LO-$a^2$ & $y^{\rm exp},a=0$ & \multicolumn{2}{c}{0.570(47)}
                              & \multicolumn{2}{c}{0.547(34)}  \\
 LO-$a^1$ & $y^{\rm exp},a=0$ & \multicolumn{2}{c}{0.519(74)}
                              &                                \\
    \bottomrule
\end{tabular}

%% file: tables/table_radsplit_1ovm.tex
%
%
\begin{tabular}{clllll} \toprule
           &     
           & \multicolumn{2}{c}{$\dmradstatminus{\rm d}$\;[\MeV]}
           & \multicolumn{2}{c}{$\dmradstatminus{\rm s}$\;[\MeV]}   \\ 
          \cmidrule(r){3-4}\cmidrule(r){5-6}
    $e$-id& $y$        & HYP1     & HYP2    & HYP1    & HYP2   \\ \midrule
       A4 & 0.0771(14) &  73(27)  &  78(21) & 133(38) & 98(26) \\
       A5 & 0.0624(13) &  71(11)  &  75( 8) &  90(11) & 76( 9) \\
       B6 & 0.0484(9)  &  70(31)  &  73(20) &  78(45) & 64(28) \\[0.15em]
       E5 & 0.0926(15) &  77(17)  &  75( 9) & 182(71) & 52(40) \\
       F6 & 0.0562(9)  &  61(31)  &  68(21) &  47(24) & 80(17) \\
       F7 & 0.0449(8)  &  82(18)  &  80(11) &  71(19) & 59(15) \\
       G8 & 0.0260(5)  &  36(26)  &  54(15) &   --    &   --   \\[0.15em]
       N5 & 0.0940(19) & 132(110) &   6(95) &   --    &   --   \\
       N6 & 0.0662(10) & 149(59)  &  94(31) &  77(31) & 76(18) \\
       O7 & 0.0447(7)  &  75(25)  &  84(17) &   --    & 67(26) \\\midrule
 LO-$a^1$ & $y^{\rm exp},a=0$ & \multicolumn{2}{c}{0.090(40)}
                              & \multicolumn{2}{c}{0.019(46)}  \\
    \bottomrule
\end{tabular}

%% file: tables/res_errors.tex
\def\sep{\hphantom{2}}
\begin{tabular}{ccccccc}
\toprule
  Source      & $\mBs$      & $\mBstar-\mB$ &  $\mBstars-\mBs$ & $\mBs-\mB$   & $\mBr-\mB$ & $\mBrs-\mBs$ \\
 $\downarrow$ & $5383(63)$  & $41.7(4.2)$   &   $37.8(3.3)$    & $88.9(5.7)$  & $791(73)$      &    $566(57)$     \\\midrule  
        A4    & \sep0.46 \% & \sep2.52 \%   & \sep3.51  \%     &    12.48  \% & \sep9.79  \%   & \sep9.93  \%      \\  
        A5    & \sep0.30 \% & \sep1.32 \%   & \sep1.81  \%     & \sep3.14  \% & \sep6.67  \%   & \sep5.78  \%      \\  
        B6    & \sep0.03 \% & \sep1.50 \%   & \sep0.39  \%     & \sep0.23  \% & \sep1.04  \%   & \sep0.06  \%      \\  
        E5    & \sep0.28 \% & \sep0.40 \%   & \sep1.80  \%     & \sep0.46  \% & \sep4.72  \%   & \sep1.51  \%      \\  
        F6    & \sep0.10 \% & \sep0.34 \%   & \sep0.45  \%     & \sep8.62  \% & \sep5.07  \%   & \sep6.72  \%      \\  
        F7    & \sep0.21 \% & \sep0.82 \%   & \sep1.50  \%     &    34.56  \% &    17.87  \%   &    23.42  \%      \\  
        G8    & \sep0.53 \% & \sep5.25 \%   & \sep0.00  \%     & \sep0.05  \% & \sep6.46  \%   & \sep0.03  \%      \\  
        N5    & \sep1.90 \% & \sep8.04 \%   & \sep0.45  \%     & \sep0.01  \% & \sep7.94  \%   & \sep0.05  \%      \\  
        N6    & \sep5.97 \% & \sep1.33 \%   &    29.18  \%     &    31.56  \% &    14.60  \%   &    27.70  \%      \\  
        O7    & \sep4.50 \% & \sep6.34 \%   &    13.49  \%     & \sep8.05  \% &    25.32  \%   &    14.42  \%      \\  
     $\omega$ &    62.84 \% &    31.75 \%   &    46.61  \%     & \sep0.04  \% & \sep0.09  \%   & \sep0.05  \%      \\  
     $\za$    &    21.13 \% & \sep0.27 \%   & \sep0.63  \%     & \sep0.75  \% & \sep0.40  \%   & \sep0.19  \%      \\  
  \bottomrule
  \end{tabular}

%% file: s4.tex
\section{Conclusions}\label{sec:conclusions}

In this paper, we have presented results for the B-meson spectrum obtained in
the framework of lattice HQET expanded to $\Or(\minv)$. Within this approach
the existence of a continuum limit is guaranteed, as numerically tested with
high accuracy in previous studies~\cite{Heitger:2004gb,Nf2tests}.  In contrast
to the HQET expansion in continuum perturbation theory, our approach is
manifestly non-perturbative in the strong coupling. We perform the continuum
extrapolation from lattice resolutions in the range 0.08--0.05\,fm. Pion
masses, in our setup with $\Nf=2$ degenerate flavours, reach down to values of
about 190\,MeV. The accuracy is at the 10\% level, for the different splittings
presented (e.g., hyperfine and $\Bd$--$\Bs$ splittings), and we always find
consistency within two standard deviations with values from the PDG, whenever a
comparison is possible.

Hyperfine splittings probe higher-order terms in HQET and the reported results
represent an important check on the validity and the reliability of the
asymptotic HQET expansion, truncated at NLO, at the b-quark mass scale.
Compared to previous quenched results, we observe a significant shift for the
hyperfine splitting in the B-meson sector, which now agrees with the
experimental determination.  In our $\Nf=2$ simulations, the hyperfine
splitting in the $\Bs$ sector appears to suffer from a residual quenching
effect, which is in line with what was seen for $\Nf=0$. In order to ascertain
that the quenching of the strange quark is indeed the root cause of the reduced
$\Bs$ hyperfine splitting seen here, we plan to extend our computations to
simulations of the $\Nf=2+1$ theory~\cite{Bruno:2014jqa}.

A dominant source of uncertainty in our results is represented by cutoff
effects (see Tab.~\ref{tab:errdist}).  This does not come unexpected since we
have not implemented $\Or(a)$ improvement at $\Or(\minv)$. While implementing a
fully non-perturbative improvement programme at this order is probably too
difficult, one may consider perturbative (tree level or one-loop) improvement
for future applications.

We also determined the mass gaps for excited states in both the B and the $\Bs$
sectors. The results are consistent with a radial splitting, e.g. as computed
for the $\Bc$ system in~\cite{Dowdall:2012ab}, but these excited states might
also be two- or multi-hadron states. 

Knowledge of the mass splittings is relevant for the computation of hadronic
parameters within the sum-rules approach and when comparing results from the
lattice to sum-rules estimates. The mass gaps of excited states are also an
important information for the computation of form-factors on the lattice, for
example for the ${\rm B} \to\pi \ell \nu$ and the $\Bs  \to \K \ell \nu$
decays, as currently endeavored by the
ALPHA~Collaboration~\cite{Bahr:2014iqa,DellaMorte:2015yda}, and in general in
the spectral analysis of two- and three-point functions.

%% file: app_ana.tex
%
\section{Determination of the statistical error in the presence of autocorrelations}
\label{sec:stat-err-auto}

In order to compute the statistical error in the presence of both
correlations between the different observables and of
autocorrelations along the HMC trajectories by which our ensembles were
generated, we employ the methods of
\cite{Wolff:2003sm,Schaefer:2010hu,ALPHAerr,ALPHAerr:LP14},
which we briefly outline below.

The starting point is the computation of the ``primary'' observables
$C^{i}_\alpha$, where $i$ labels the $N_{\rm meas}$ gauge configurations,
and $\alpha$ is an aggregate label for the different correlators
measured (stat, spin, kin), the Euclidean time separation $t$ between
the source and sink, and the different smearing levels employed at
source and sink.
The gauge average $\overline{C}_\alpha$ and the variance
$\sigma^2_{C_\alpha}$ are computed as usual.
To estimate the true statistical error $\sigma_{\overline{C}_\alpha}$
of the gauge average, we also require the integrated autocorrelation
time $\tau^{\rm int}_{C_\alpha}$, which is computed from the
autocorrelation function
\begin{equation}
  \Gamma^{(1)}_{\alpha\beta}\equiv\Gamma_{C_{\alpha} C_{\beta}}(\tau)
   = \lim_{K\to\infty}\frac{1}{K}\sum_{i=1}^K
     \left(C_\alpha^{i+\tau}-\overline{C}_\alpha\right)\left(C_\beta^{i}-\overline{C}_\beta\right)\,,
\end{equation} 
where $\tau$ is the separation in simulation time along the Markov chain.
In order to take into account the long-time tail of $\Gamma$, the
conservative estimate
\cite{Schaefer:2010hu}
\begin{equation}\label{eq:tauintest}
  \tau^{\rm int}_{C_\alpha} = \frac{1}{2}+\frac{1}{\Gamma^{(1)}_{\alpha\alpha}(0)}
  \left(\sum_{\tau=1}^{W-1}\Gamma^{(1)}_{\alpha\alpha}(\tau) +\tau^{\rm exp}\Gamma^{(1)}_{\alpha\alpha}(W)\right)
\end{equation}
is used, where $\tau^{\rm exp}$ is an estimate of the exponential
autocorrelation time of the Markov chain. The values used in our analysis are
listed in Table~\ref{tab:meas-parms}.  
The window size $W$ is automatically chosen as the point $\tau=W$ where
$\Gamma_{\alpha\alpha}(\tau)$ comes close to zero within about 1.5 of its
estimated error. The true statistical error is then given by 
\begin{equation}
  \sigma^2_{\overline{C}_\alpha} = 2 \tau^{\rm int}_{C_\alpha} \frac{\sigma^2_{C_\alpha}}{N_{\rm meas}} \,.
\end{equation}

For derived observables $D_{\alpha'}$, which are functions of gauge averages of
the primary observables $\overline{C}_\alpha$, and in our case include the
generalized eigenvalues and eigenvectors as well as the energies derived from
them, we compute the derivatives
\begin{equation}
  J_{\alpha'\alpha} \equiv \frac{\partial D_{\alpha'}}{\partial\overline{C}_\alpha}
\end{equation}
and the autocorrelation function
\cite{Schaefer:2010hu}
\begin{equation}
  \Gamma^{(2)}_{\alpha'\beta'}(\tau)\equiv\sum_{\alpha,\beta} J_{\alpha'\alpha} \Gamma^{(1)}_{\alpha\beta}(\tau) J_{\beta'\beta} \,.
\end{equation} 
The variance of the derived observable $D_{\alpha'}$ is then given by
\begin{equation}
  \sigma^2_{D_{\alpha'}} = \Gamma^{(2)}_{\alpha'\alpha'}(0)
\end{equation}
and its statistical error by
\begin{equation}
  \label{eq:dererrorest}
  \sigma^2_{\overline{D}_\alpha'} = 2 \tau^{\rm int}_{D_\alpha'} \frac{\sigma^2_{D_\alpha'}}{N_{\rm meas}} \,,
\end{equation}
where the integrated autocorrelation time $\tau^{\rm int}_{D_\alpha'}$ is again
estimated using (\ref{eq:tauintest}), with $\Gamma^{(2)}$ substituted for
$\Gamma^{(1)}$.

Since the extraction of plateau values from a weighted fit requires knowledge
of the errors of the individual points, in principle this procedure should be
iterated, with the plateau averages as secondary derived observables of the
derived observables. However, the integrated autocorrelation times for the
effective energies in the plateau region do not markedly differ. Therefor, it
is sufficient to employ their variances (as estimated using e.g. a Jackknife
procedure for error propagation) to weight the fit, treating only the fitted
values as derived observables. This simplified procedure has been adopted here.

In order to extract the final answer in physical units, the plateau values must
be combined with each other and with the HQET parameters and lattice spacing,
propagating the errors on each of those to the final result, where the HQET
parameters are statistically independent of the large-volume observables;
similarly, in extrapolating to the chiral and continuum limits, the results
obtained from different ensembles are statistically independent, and their
contribution to the error of the final result $f$ can thus be added in
quadrature:
\begin{equation}
  \sigma^2_f = \sum_e \sigma^2_f(e) + \sum_{i,j} \frac{\partial f}{\partial Y_i}C_{Y_iY_j}\frac{\partial f}{\partial Y_j} \,,
\end{equation}
where the $Y_i$ are the additional parameters, $C_{Y_iY_j}$ is their (known)
covariance matrix, and $\sigma^2_f(e)$ is the error computed according to
eqn.~(\ref{eq:dererrorest}) when taking into account only the fluctuations of
$C_\alpha$ on ensemble $e$~\cite{Schaefer:2010hu,ALPHAerr,ALPHAerr:LP14}.

%

%% file: app_Eeff.tex
%
%
\section{Effective energies and matrix elements}

In this Appendix we provide the numerical results of our GEVP analysis after
performing a weighted plateaux average
\begin{align}\label{eq:plateau_Ex}
    E^{\rm x}_{n} &= \frac{ \sum_{t} w(t) E^{\rm eff,x}_{n}(t,t_0) }{ \sum_{t} w(t) }  \;, & 
             w(t) &= \left( \sigma \big[E^{\rm eff,x}_{n}(t,t_0) \big]\right)^{-2}
\end{align}
for $t \in [\tmin,\tmax]$ and $t_0\geq\tmin/2$, see Section~\ref{s:gevp}. The
specific value of $\tmax$ is irrelevant due to noise dominated data at large
time separations.%
\footnote{For plotting convenience we set $\tmax/a=\infty$ in the plots.}
As a consequence of the exponential growth of the noise-to-signal ratio, the
quoted errors are dominated by the error at $\tmin$ and we decided to quote
$\tmin$ as subscript to the statistical error in the following Tables. For
determining $\tmin$, we use $t_0=t-1$ in the GEVP, and then $t_0 = \tmin-1$
once $\tmin$ is fixed.  The errors quoted are those entering the corresponding
effective energy plots presented below and do not contain the tail contribution
of the error.  
Since the autocorrelation is expected to be the same among different time
slices in the plateau region, it does not affect the
estimate~\eqref{eq:plateau_Ex} and can be added, including the exponential
tail, whenever needed explicitly. In fact, all values quoted for derived
observables as presented in the main text have the exponential tail included as
discussed in Appendix~\ref{sec:stat-err-auto}.

The same procedure has been used to obtain matrix elements $p^{\rm x}$. We
include them in the present paper (Table~\ref{tab:hl_peff0}-\ref{tab:hs_peff0})
in order to complete the data set used
in~\cite{Bernardoni:2013xba,Bernardoni:2014fva}.

A special case is the ensemble labeled A5 for which we have two independent
Monte Carlo histories/replicas (A5c and A5d). Here, measurements and
a subsequent GEVP analysis have been independently performed for both replicas
in the heavy-light sector but only on A5d for heavy-strange.  The results from
different replicas are then combined before derived observables as presented in
the main text are computed.

%
%

\begin{table}[t]
  \centering\small
  \input{tables/table_Eeff0_hl.tex}

  \caption{Raw data for plateau-averaged ground state energies in the heavy-light sector.
           The subscript to the statistical error is the value of $\tmin$ in the GEVP analysis.}
  \label{tab:hl_Eeff0}
\end{table}

\begin{table}[t]
  \centering\small
  \input{tables/table_Eeff1_hl.tex}

  \caption{Raw data for plateau-averaged 1$^{\rm st}$ excited state energies in the heavy-light sector.
           The subscript to the statistical error is the value of $\tmin$ in the GEVP analysis.}
  \label{tab:hl_Eeff1}
\end{table}

\begin{table}[t]
  \centering\small
  \input{tables/table_Eeff0_hs.tex}

  \caption{Raw data for plateau-averaged ground state energies in the heavy-strange sector.
           The subscript to the statistical error is the value of $\tmin$ in the GEVP analysis.}
  \label{tab:hs_Eeff0}
\end{table}

\begin{table}[t]
  \centering\small
  \input{tables/table_Eeff1_hs.tex}

  \caption{Raw data for plateau-averaged 1$^{\rm st}$ excited state energies in the heavy-strange sector.
           The subscript to the statistical error is the value of $\tmin$ in the GEVP analysis.}
  \label{tab:hs_Eeff1}
\end{table}

\begin{table}[t]
  \centering\small
  \input{tables/table_peff0_hl.tex}

  \caption{Raw data for plateau-averaged ground state matrix elements in the heavy-light sector.
           The subscript to the statistical error is the value of $\tmin$ in the GEVP analysis.}
  \label{tab:hl_peff0}
\end{table}

\begin{table}[t]
  \centering\small
  \input{tables/table_peff0_hs.tex}

  \caption{Raw data for plateau-averaged ground state matrix elements in the heavy-strange sector.}
  \label{tab:hs_peff0}
\end{table}

%
%

\begin{figure}[htb]
   \centering
   \includegraphics[width=0.9\textwidth]{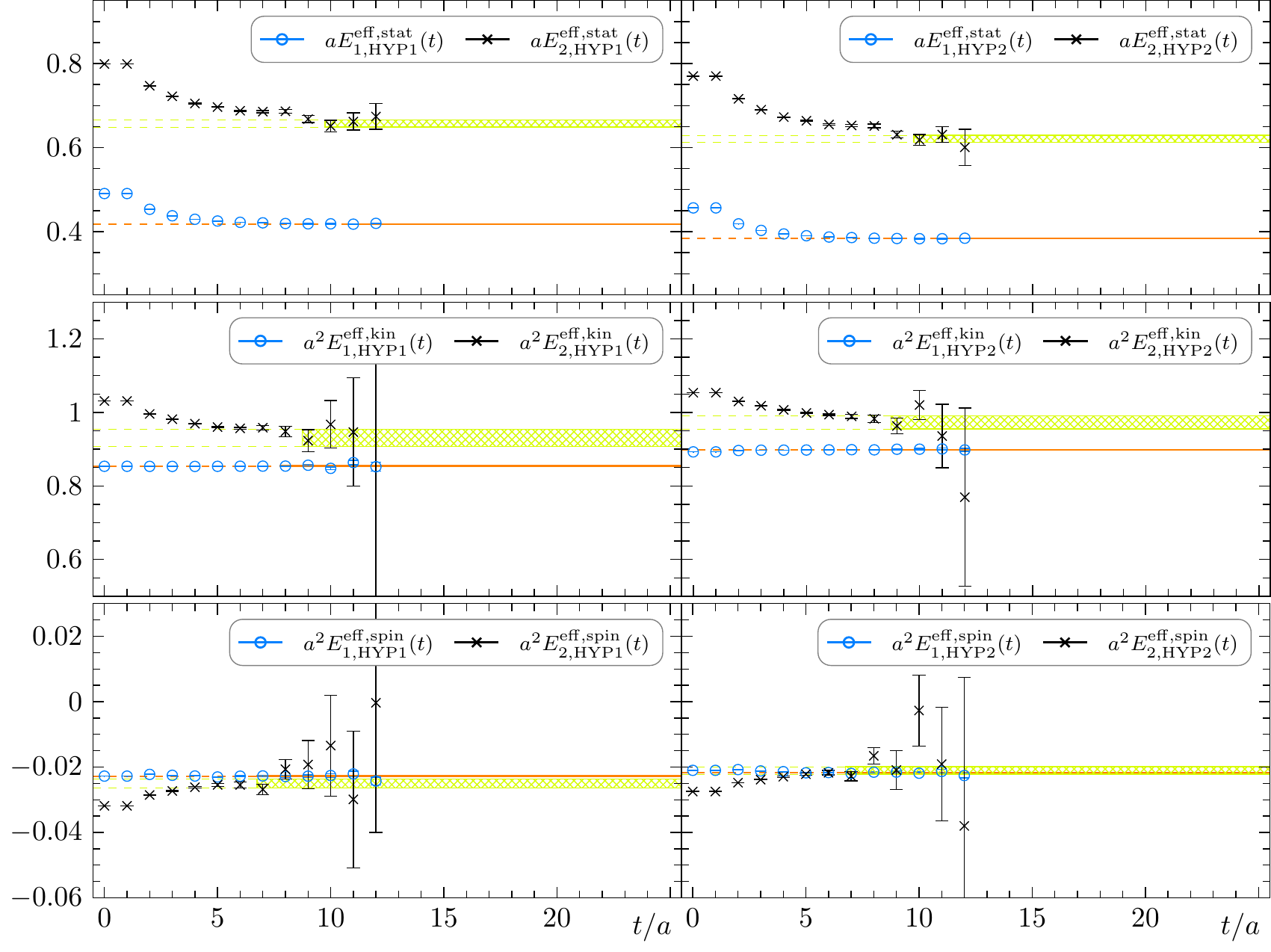}
   \vskip1em
   \includegraphics[width=0.9\textwidth]{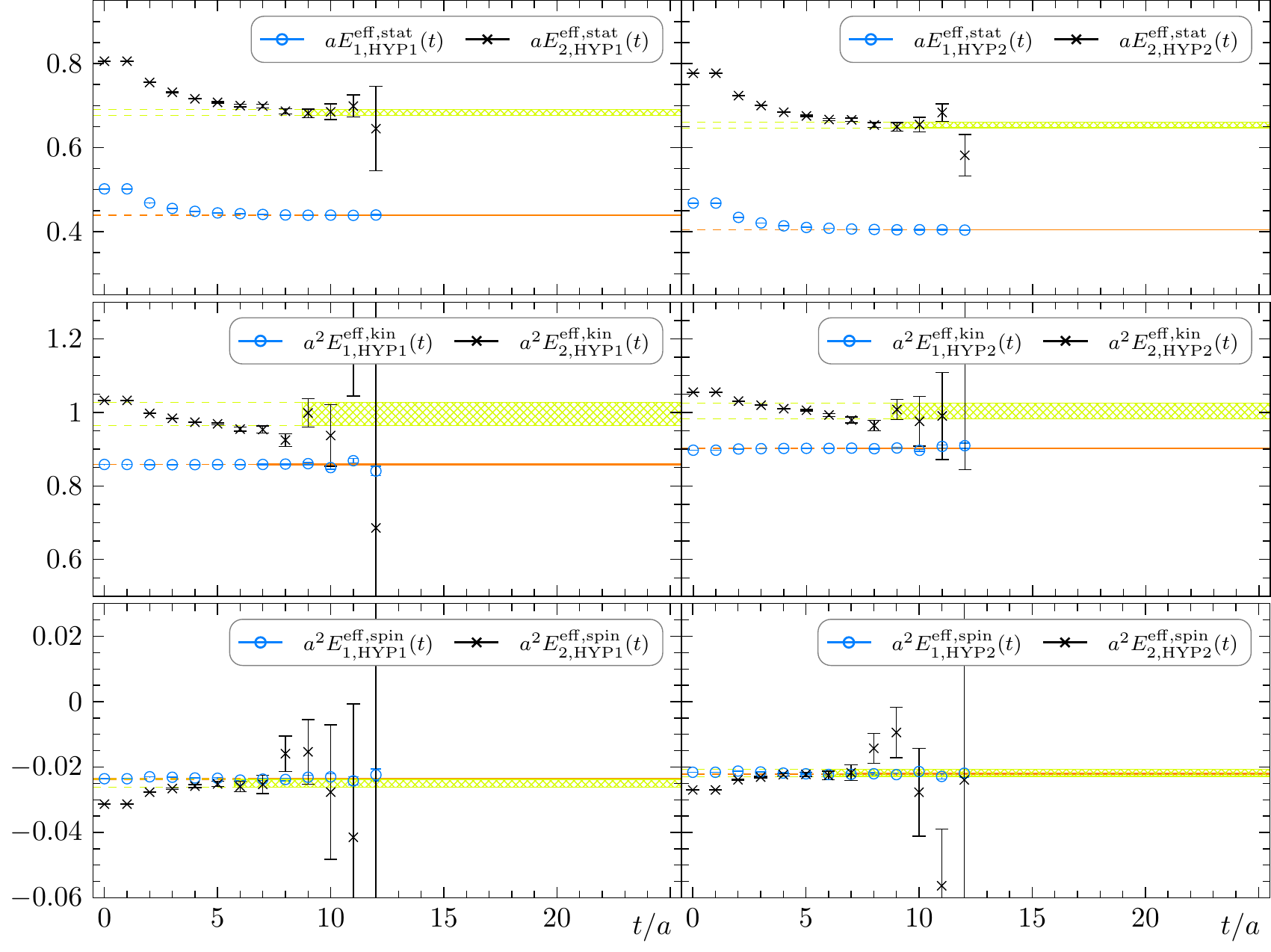}
   \caption{Effective energies $\Exnd$ following our GEVP analysis in the 
            heavy-light (\emph{top}) and heavy-strange 
            (\emph{bottom}) sector on ensemble A4.
           }
   \label{fig:Ex_A4}
\end{figure}

\begin{figure}[htb]
   \centering
   \includegraphics[width=0.9\textwidth]{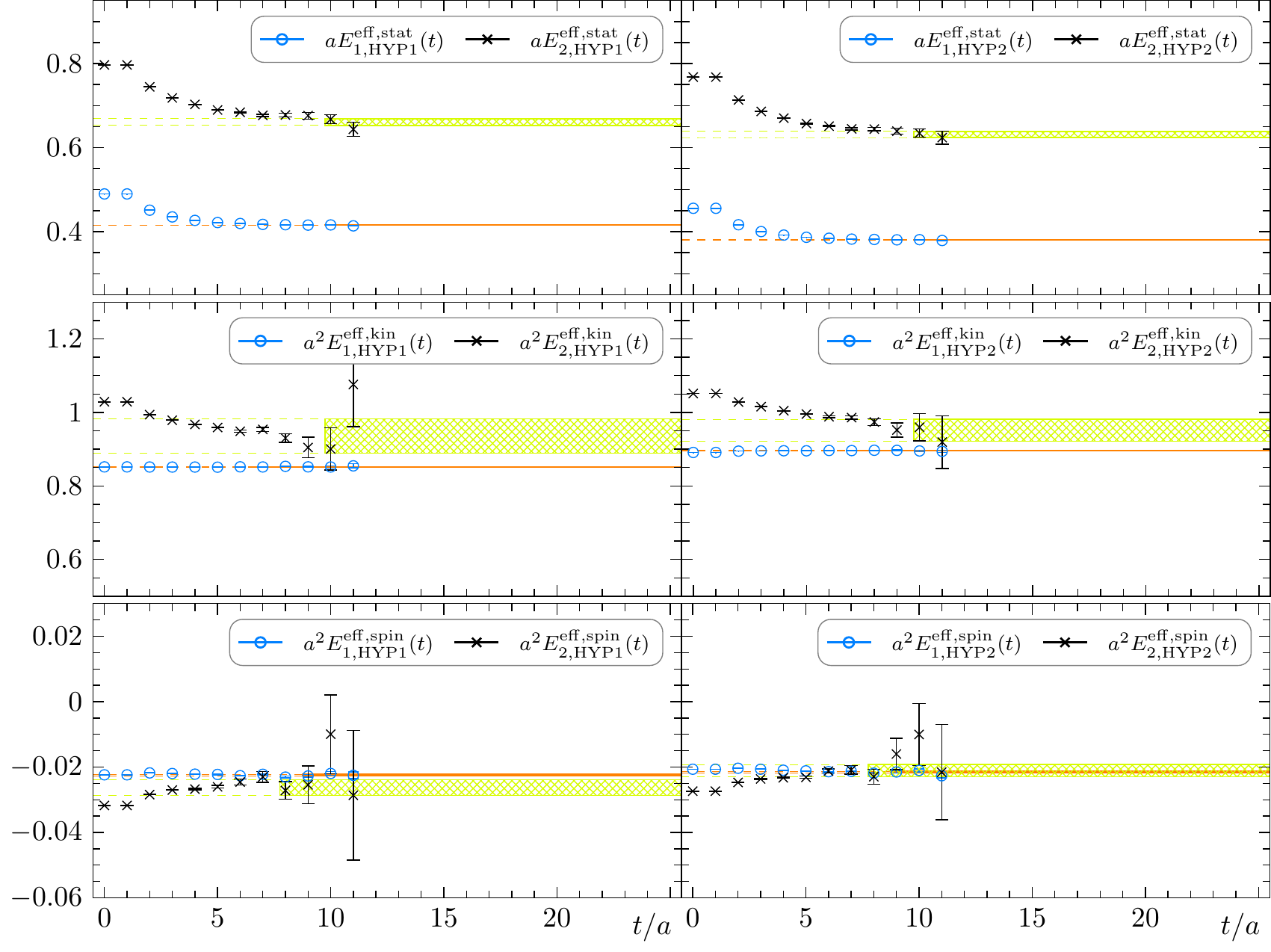}
   \vskip1em
   \includegraphics[width=0.9\textwidth]{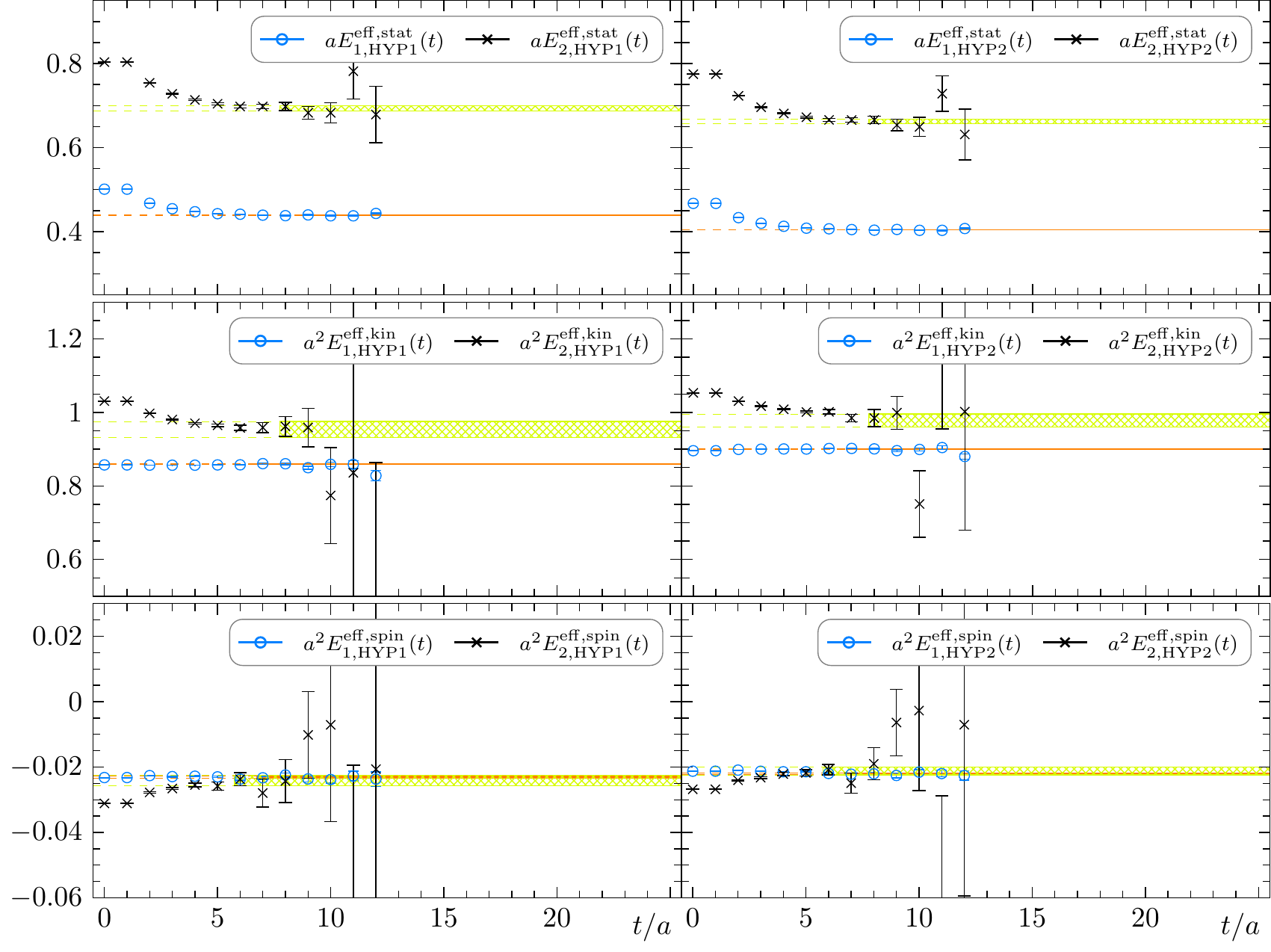}
   \caption{Effective energies $\Exnd$ following our GEVP analysis in the 
            heavy-light (\emph{top}) and heavy-strange 
            (\emph{bottom}) sector on ensemble A5c.
           }
   \label{fig:Ex_A5c}
\end{figure}

\begin{figure}[htb]
   \centering
   \includegraphics[width=0.9\textwidth]{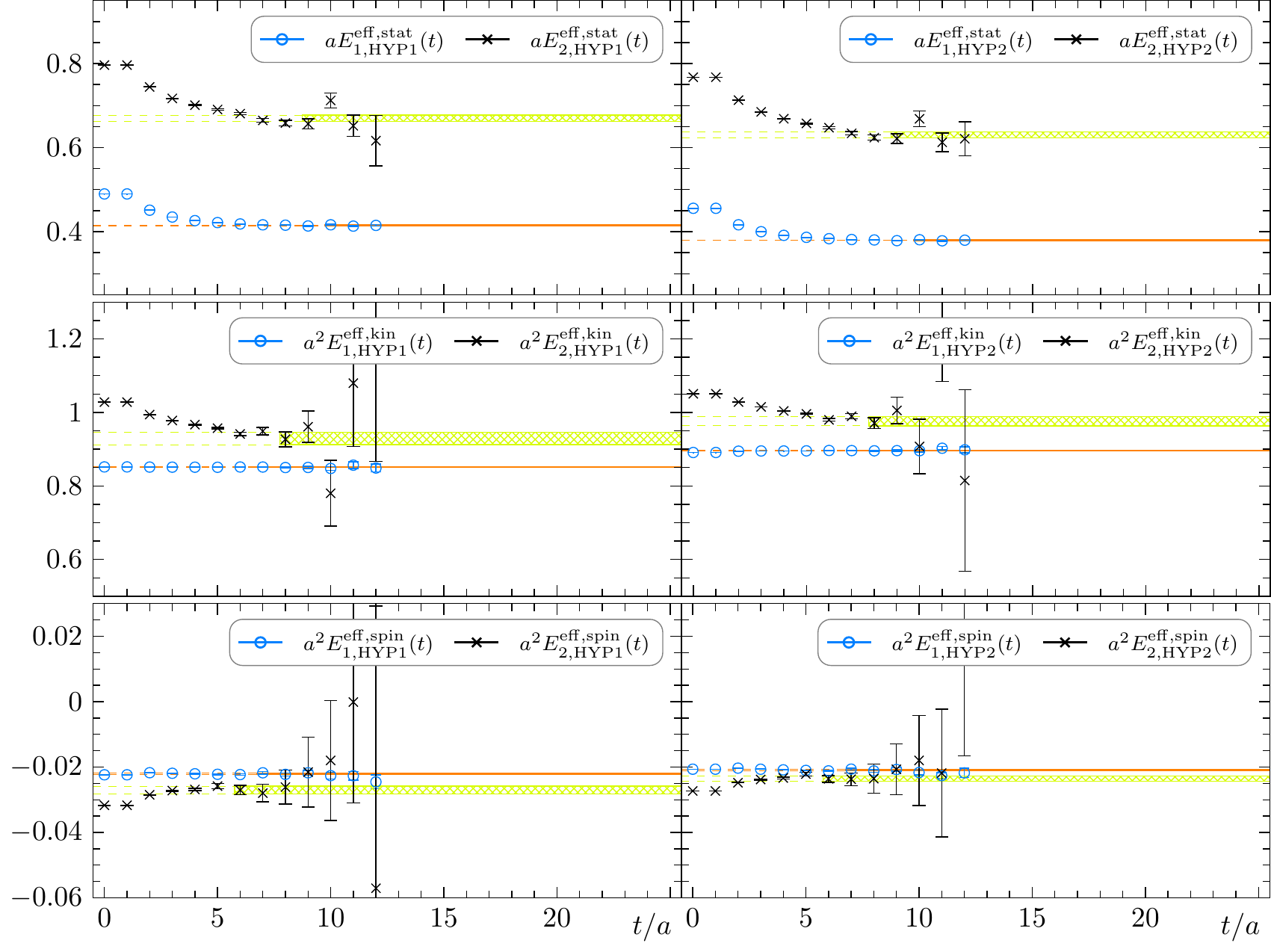}
   \caption{Effective energies $\Exnd$ following our GEVP analysis in the 
            heavy-strange sector on ensemble A5d.
           }
   \label{fig:Ex_A5d}
\end{figure}

\begin{figure}[htb]
   \centering
   \includegraphics[width=0.9\textwidth]{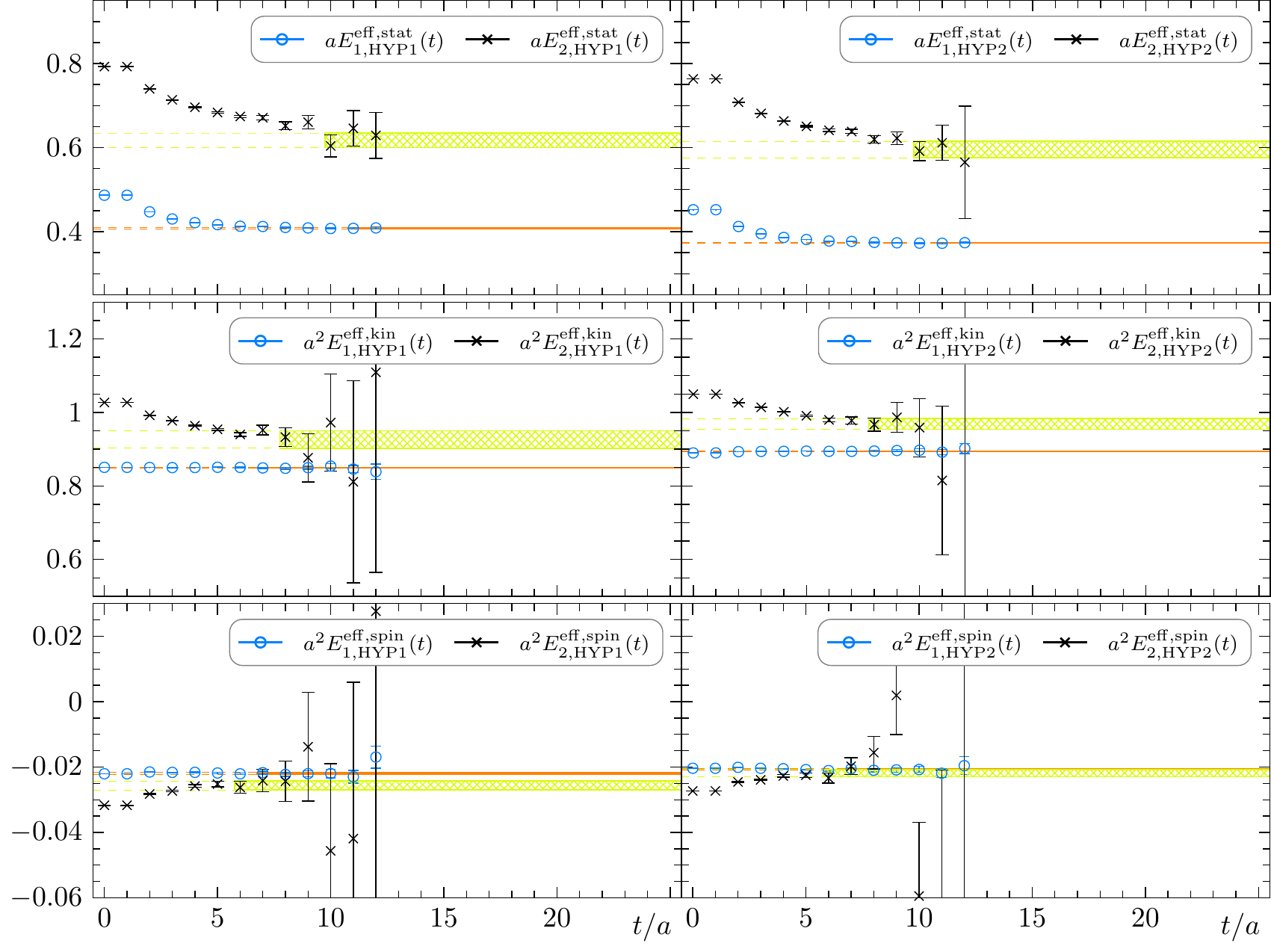}
   \vskip1em
   \includegraphics[width=0.9\textwidth]{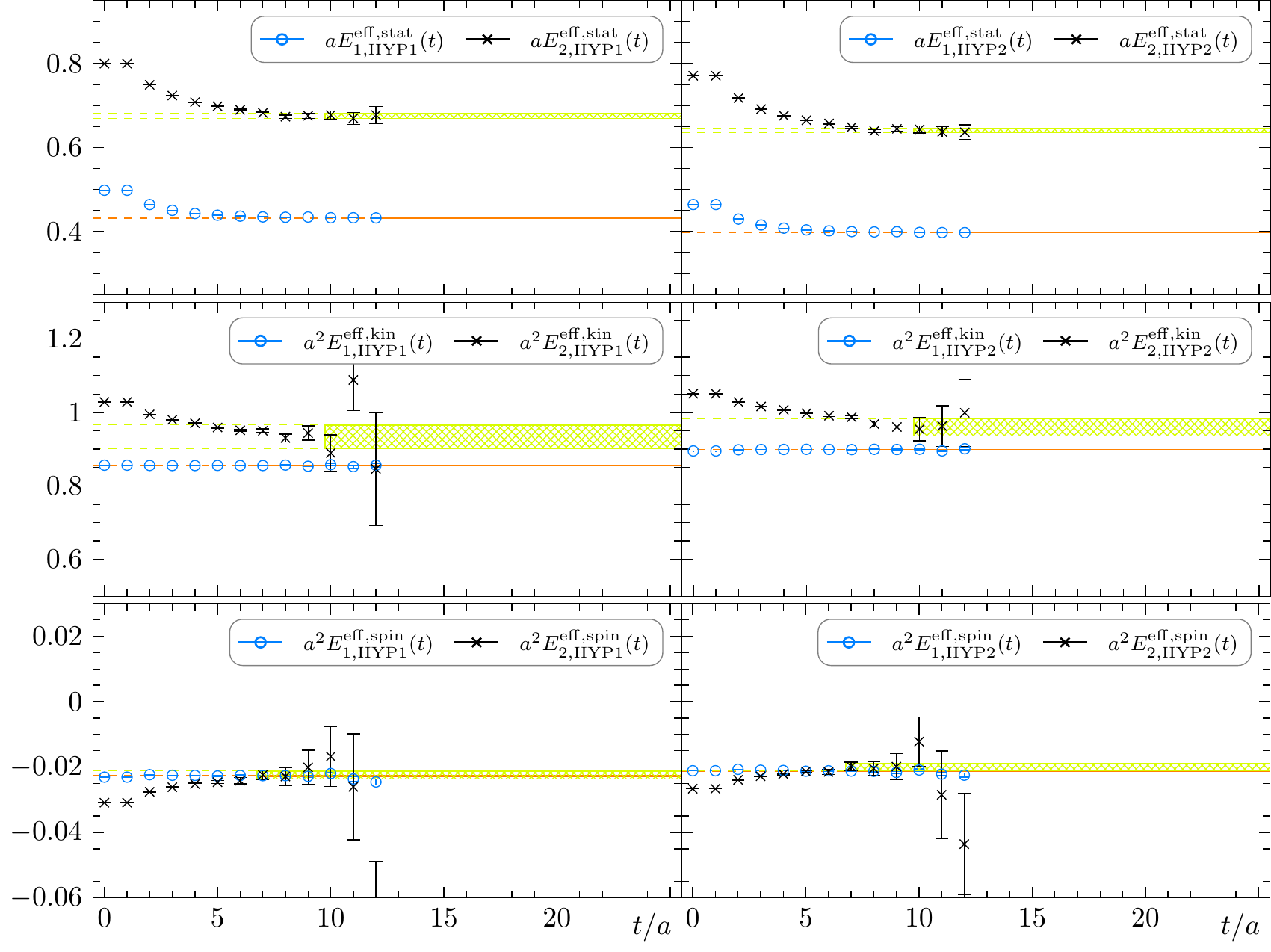}
   \caption{Effective energies $\Exnd$ following our GEVP analysis in the 
            heavy-light (\emph{top}) and heavy-strange 
            (\emph{bottom}) sector on ensemble B6.
           }
   \label{fig:Ex_B6}
\end{figure}

\begin{figure}[htb]
   \centering
   \includegraphics[width=0.9\textwidth]{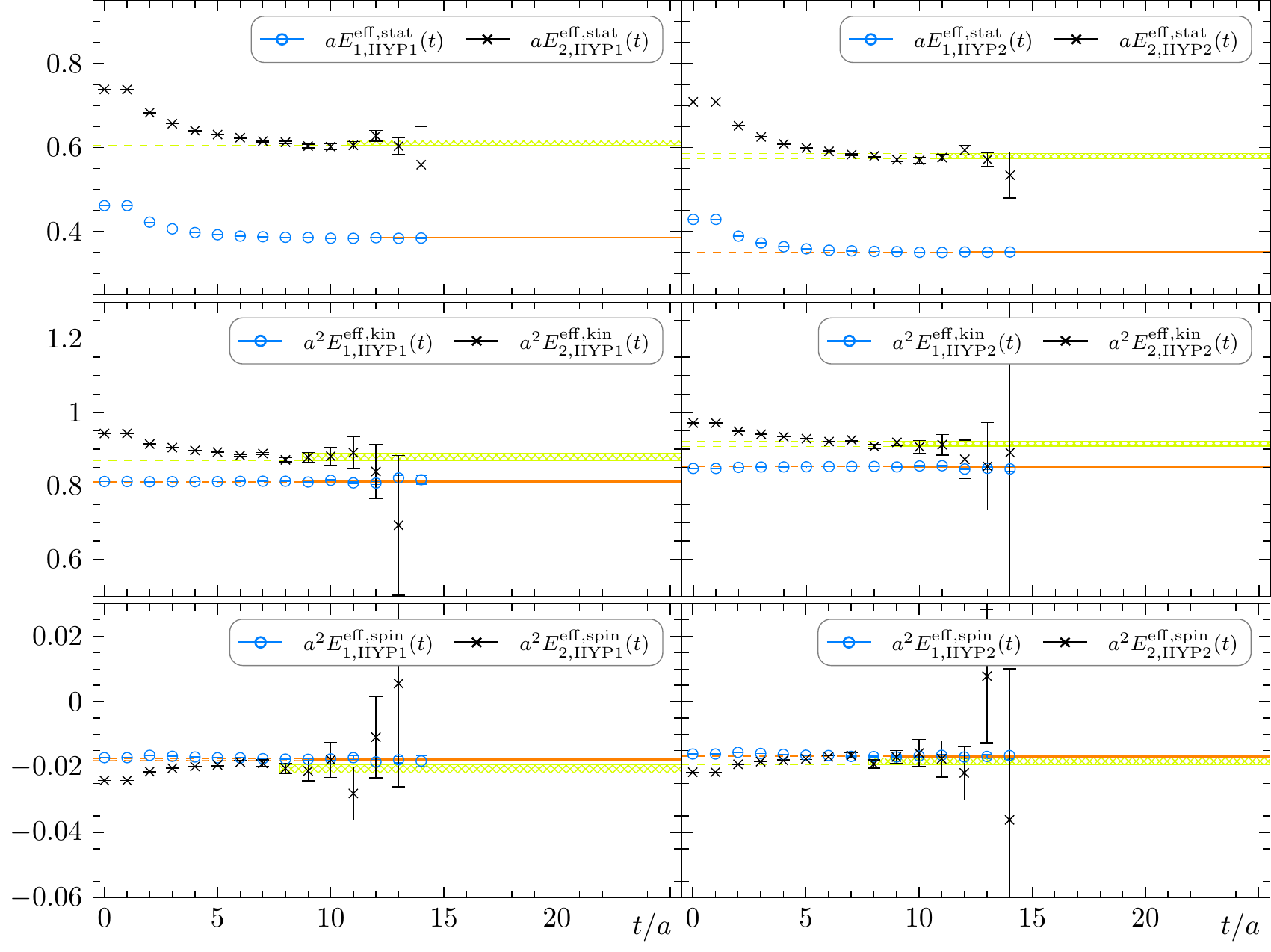}
   \vskip1em
   \includegraphics[width=0.9\textwidth]{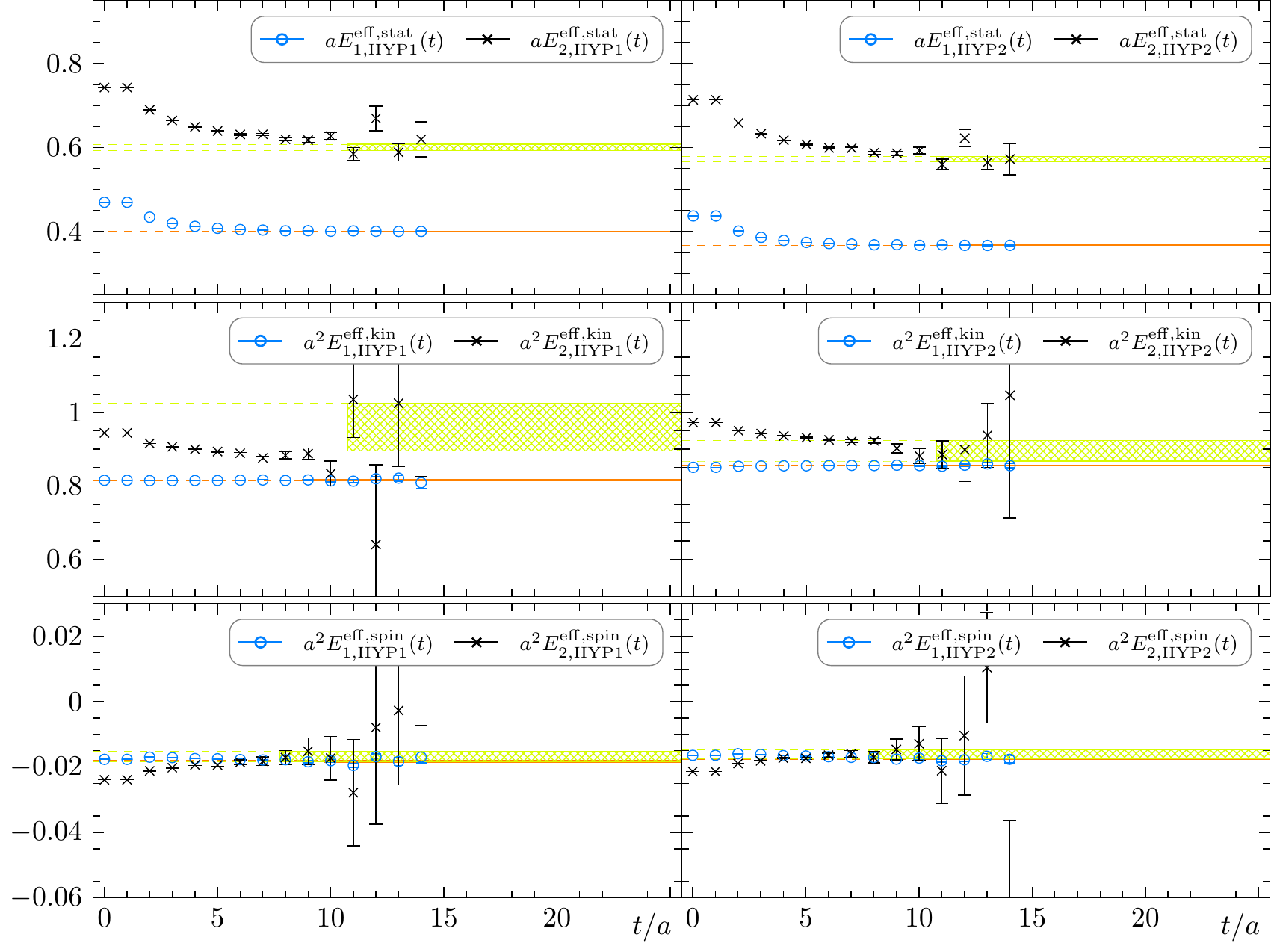}
   \caption{Effective energies $\Exnd$ following our GEVP analysis in the 
            heavy-light (\emph{top}) and heavy-strange 
            (\emph{bottom}) sector on ensemble E5g.
           }
   \label{fig:Ex_E5g}
\end{figure}

\begin{figure}[htb]
   \centering
   \includegraphics[width=0.9\textwidth]{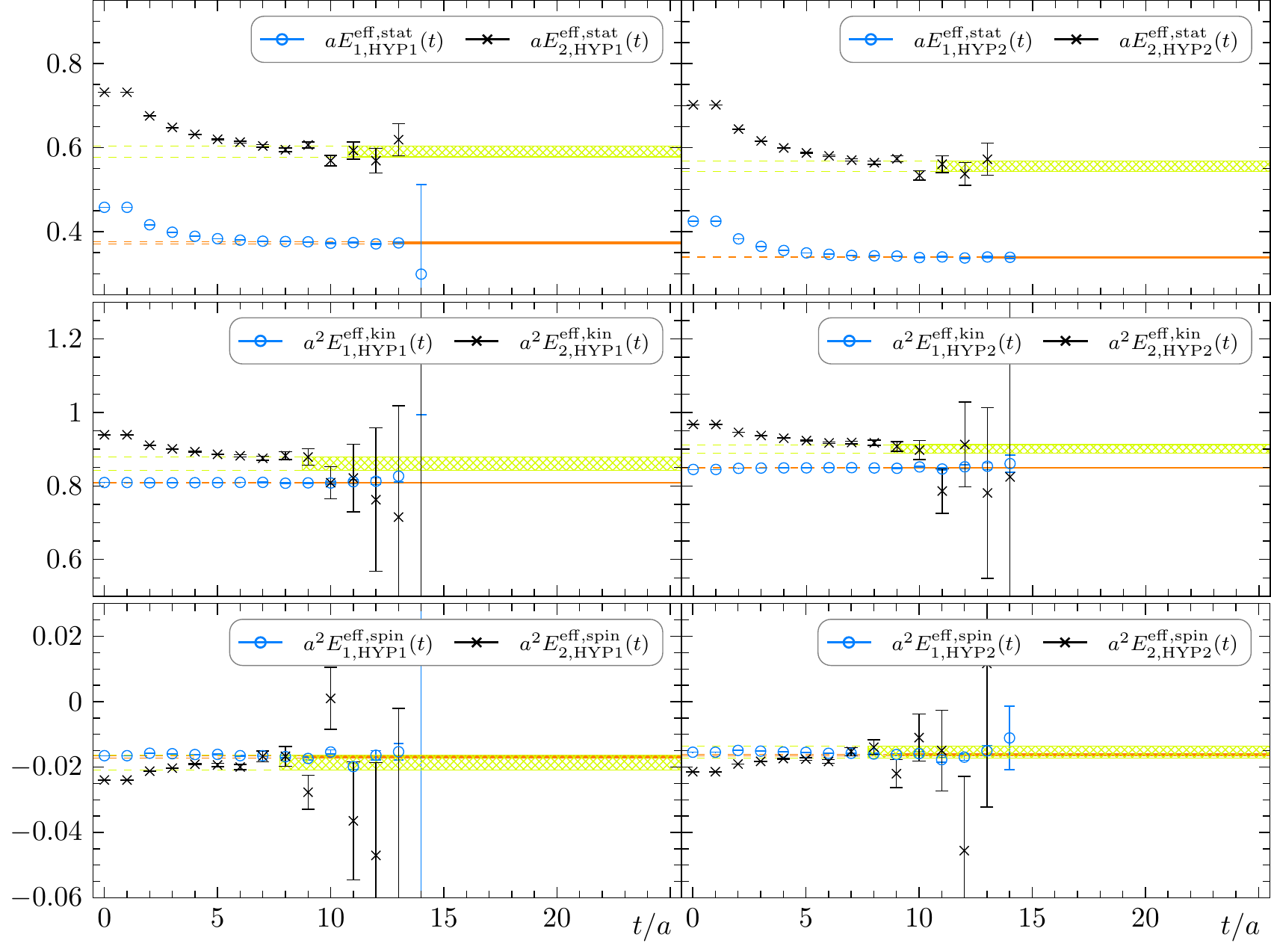}
   \vskip1em
   \includegraphics[width=0.9\textwidth]{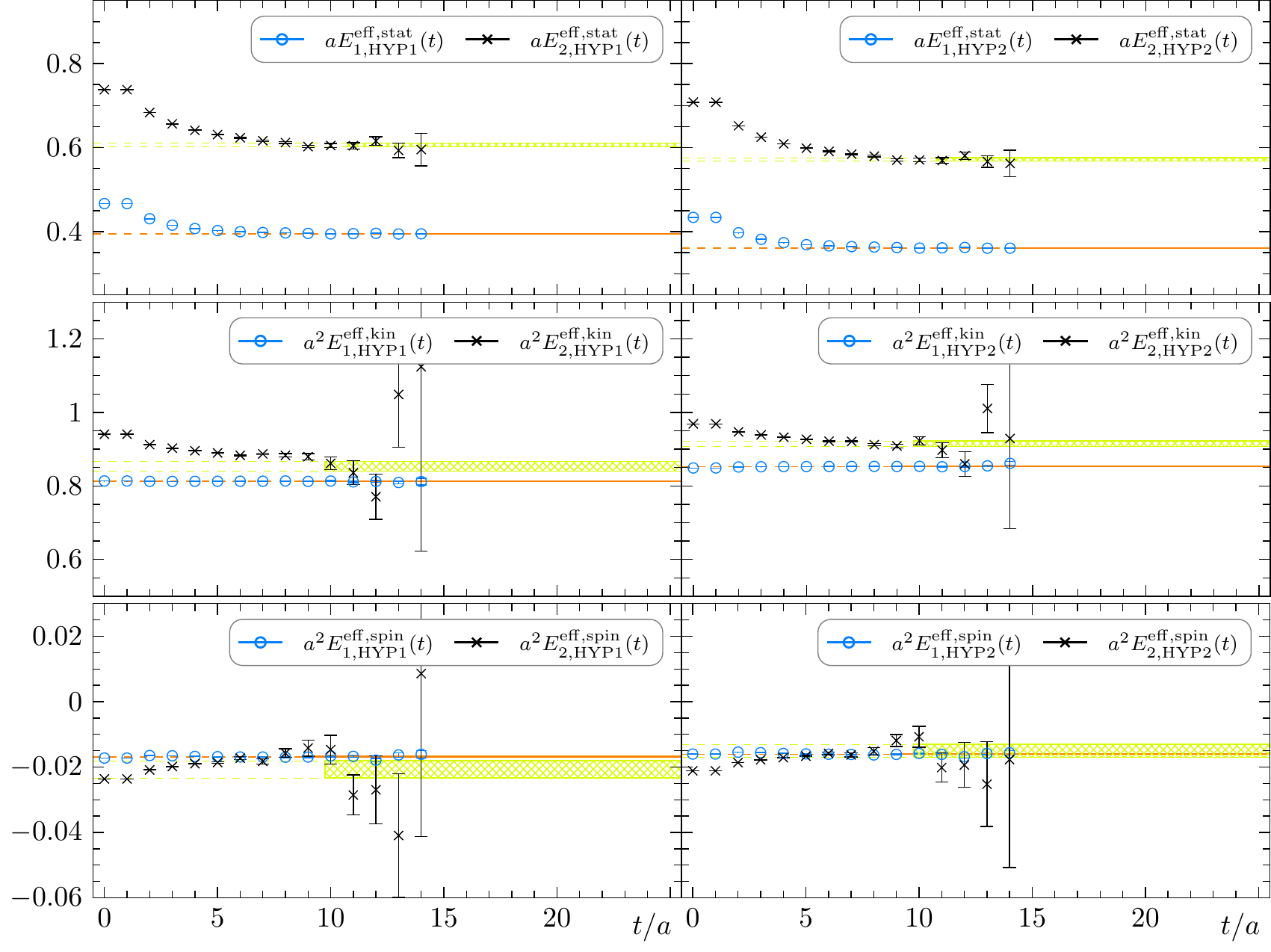}
   \caption{Effective energies $\Exnd$ following our GEVP analysis in the 
            heavy-light (\emph{top}) and heavy-strange 
            (\emph{bottom}) sector on ensemble F6.
           }
   \label{fig:Ex_F6}
\end{figure}

\begin{figure}[htb]
   \centering
   \includegraphics[width=0.9\textwidth]{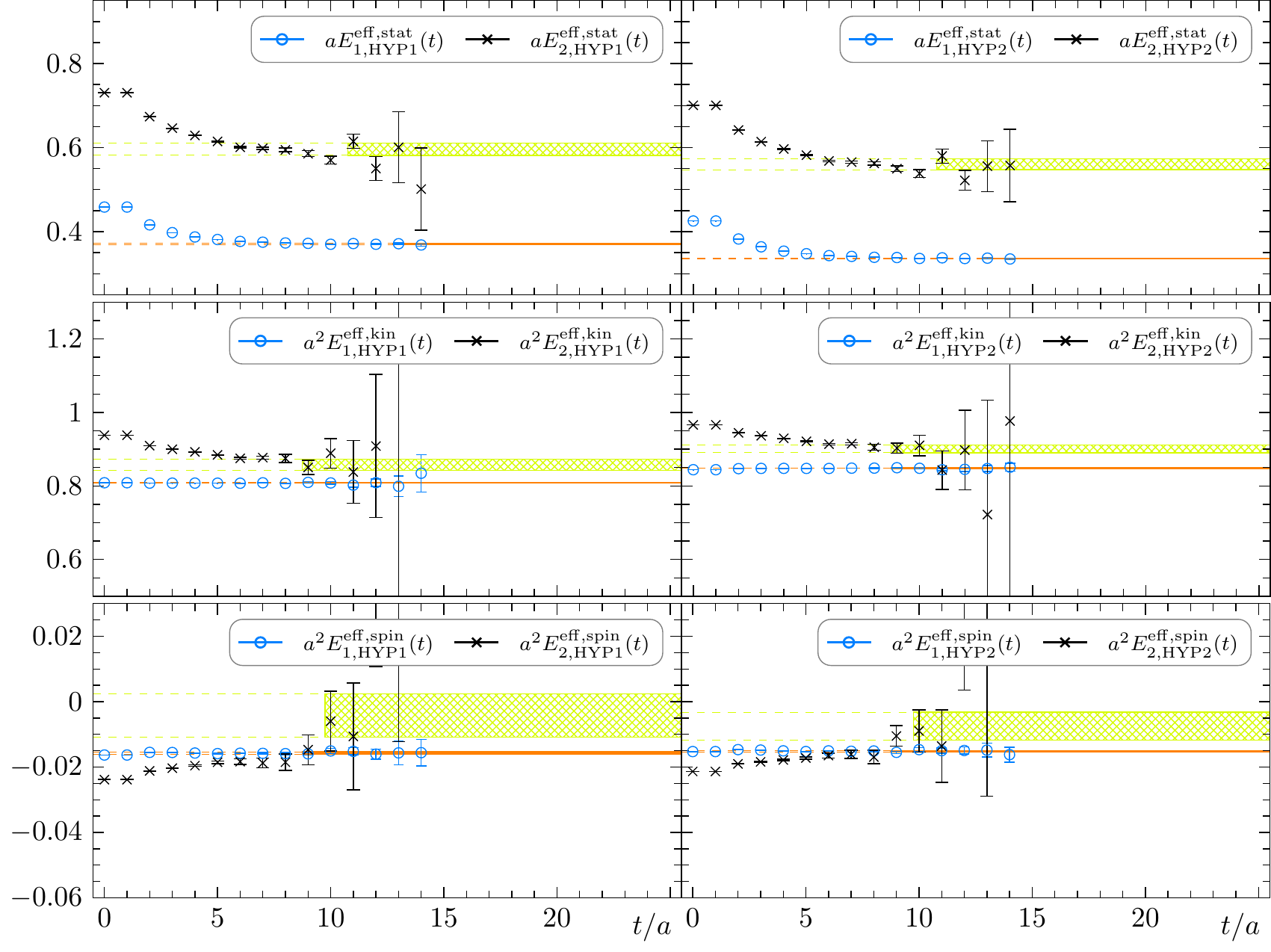}
   \vskip1em
   \includegraphics[width=0.9\textwidth]{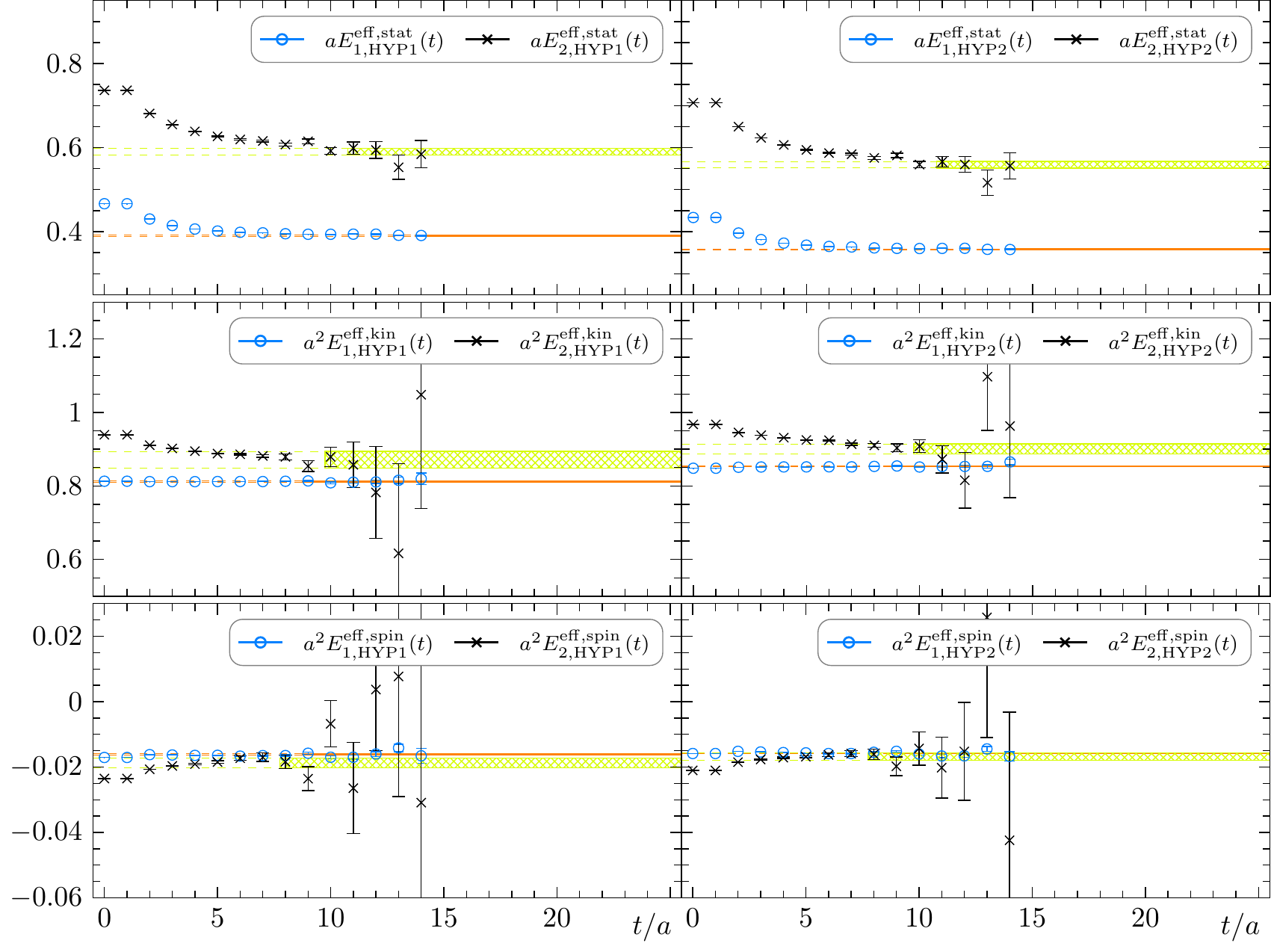}
   \caption{Effective energies $\Exnd$ following GEVP analysis in the 
            heavy-light (\emph{top}) and heavy-strange 
            (\emph{bottom}) sector on ensemble F7.
           }
   \label{fig:Ex_F7}
\end{figure}

\begin{figure}[htb]
   \centering
   \includegraphics[width=0.9\textwidth]{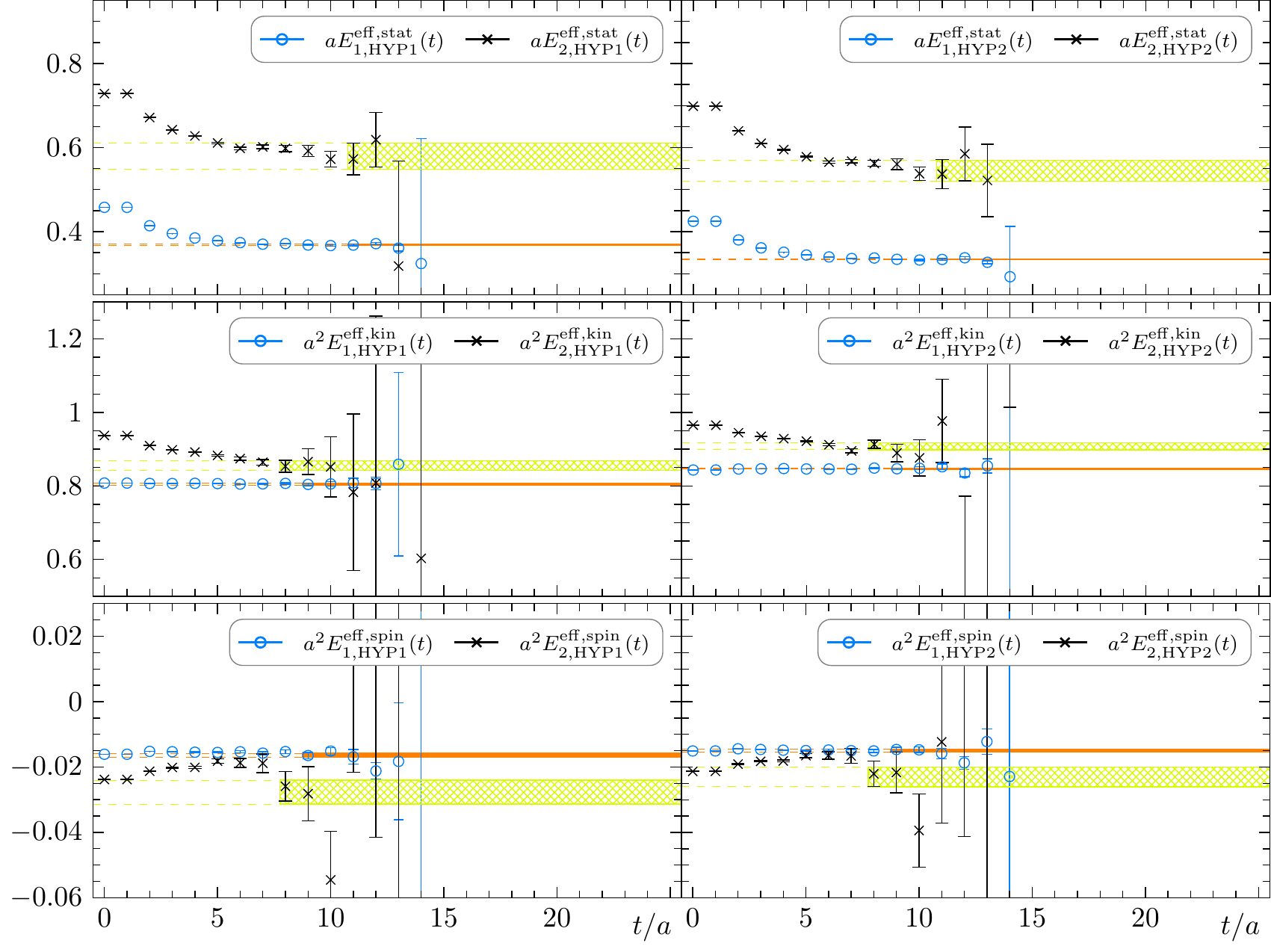}
   \caption{Effective energies $\Exnd$ following our GEVP analysis in the 
            heavy-light sector on ensemble G8.
           }
   \label{fig:Ex_G8}
\end{figure}

\begin{figure}[htb]
   \centering
   \includegraphics[width=0.9\textwidth]{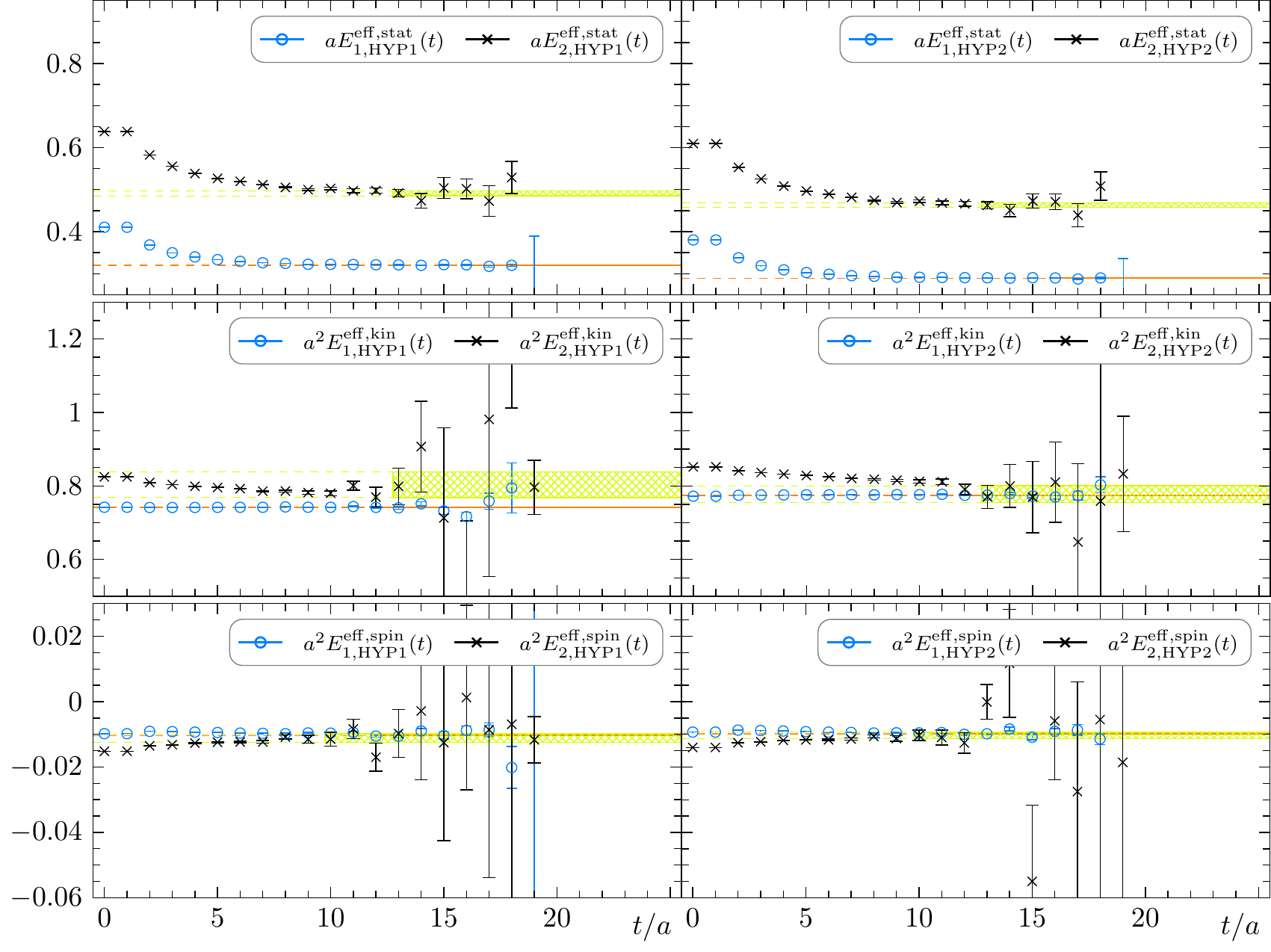}
   \caption{Effective energies $\Exnd$ following our GEVP analysis in the 
            heavy-light sector on ensemble N5.
           }
   \label{fig:Ex_N5}
\end{figure}

\begin{figure}[htb]
   \centering
   \includegraphics[width=0.9\textwidth]{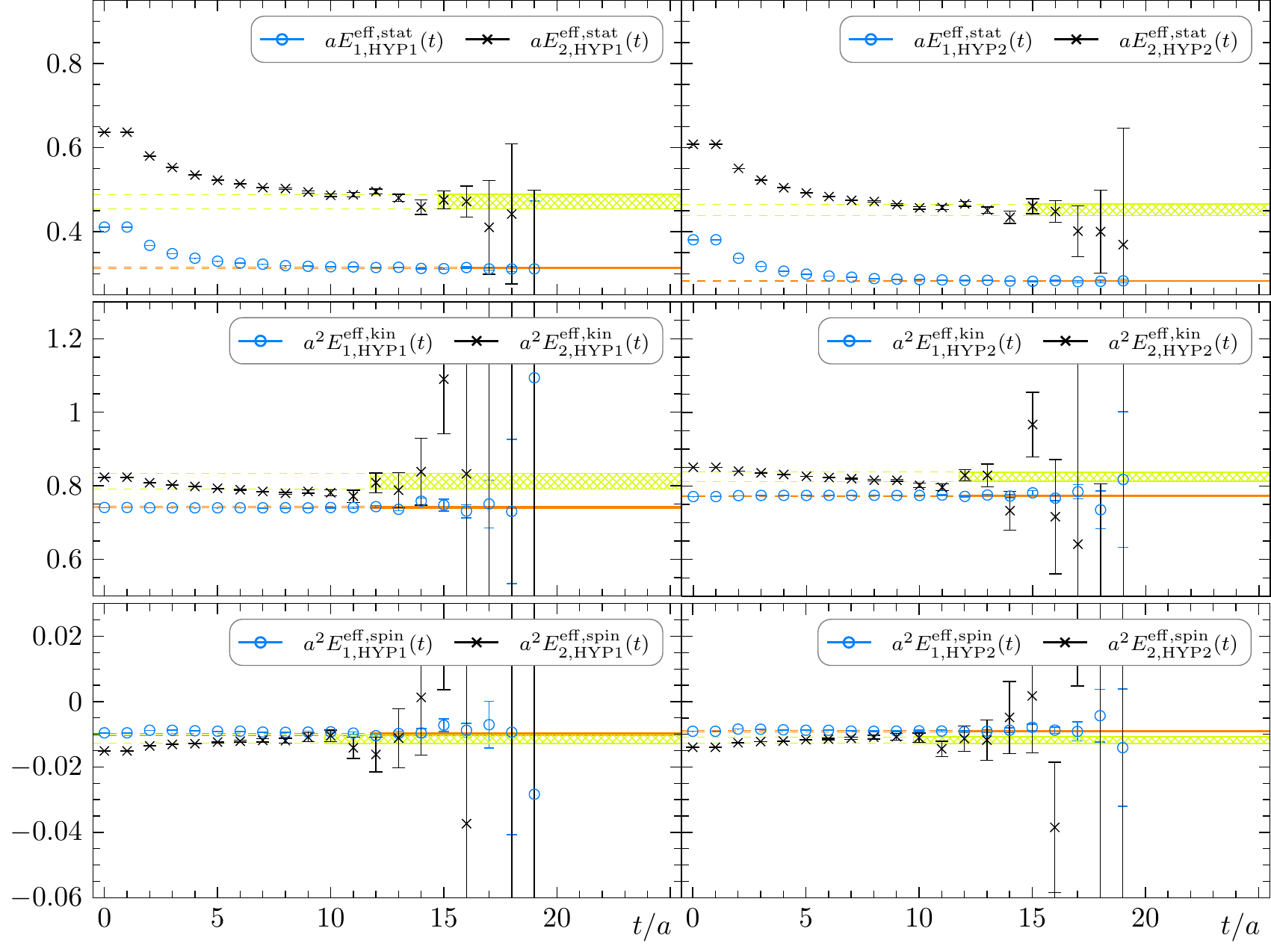}
   \vskip1em
   \includegraphics[width=0.9\textwidth]{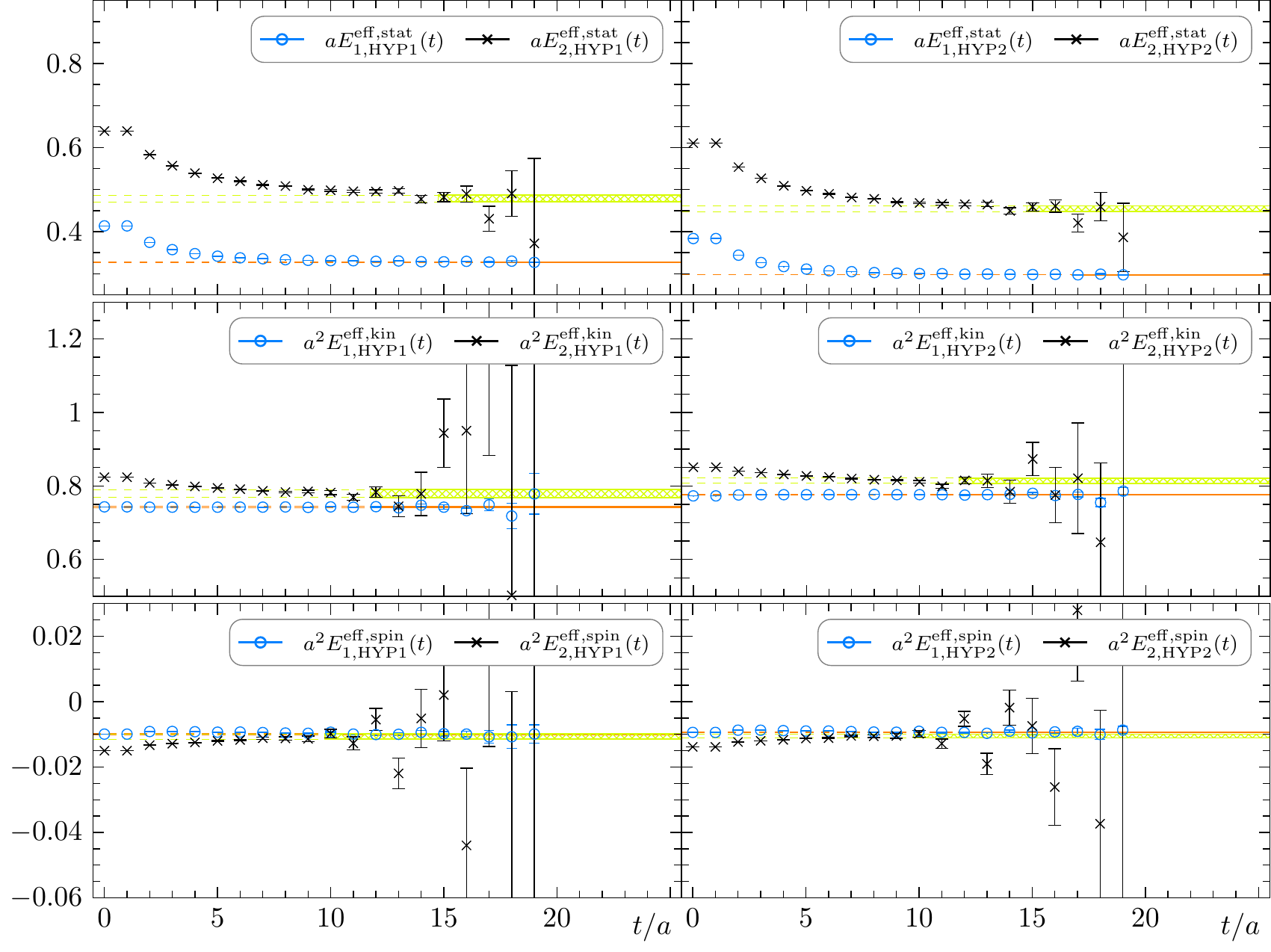}
   \caption{Effective energies $\Exnd$ following our GEVP analysis in the 
            heavy-light (\emph{top}) and heavy-strange 
            (\emph{bottom}) sector on ensemble N6.
           }
   \label{fig:Ex_N6}
\end{figure}

\begin{figure}[htb]
   \centering
   \includegraphics[width=0.9\textwidth]{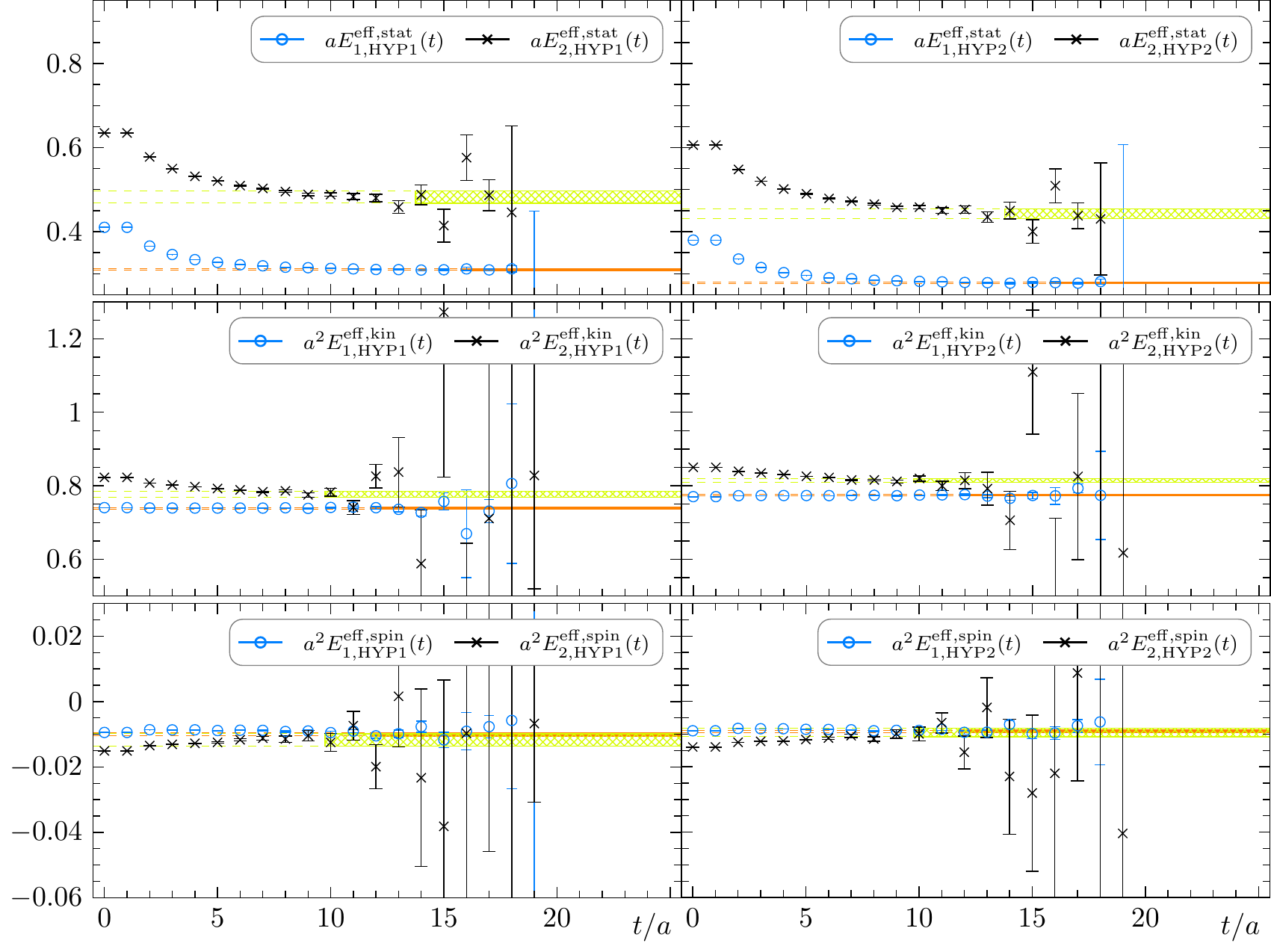}
   \vskip1em
   \includegraphics[width=0.9\textwidth]{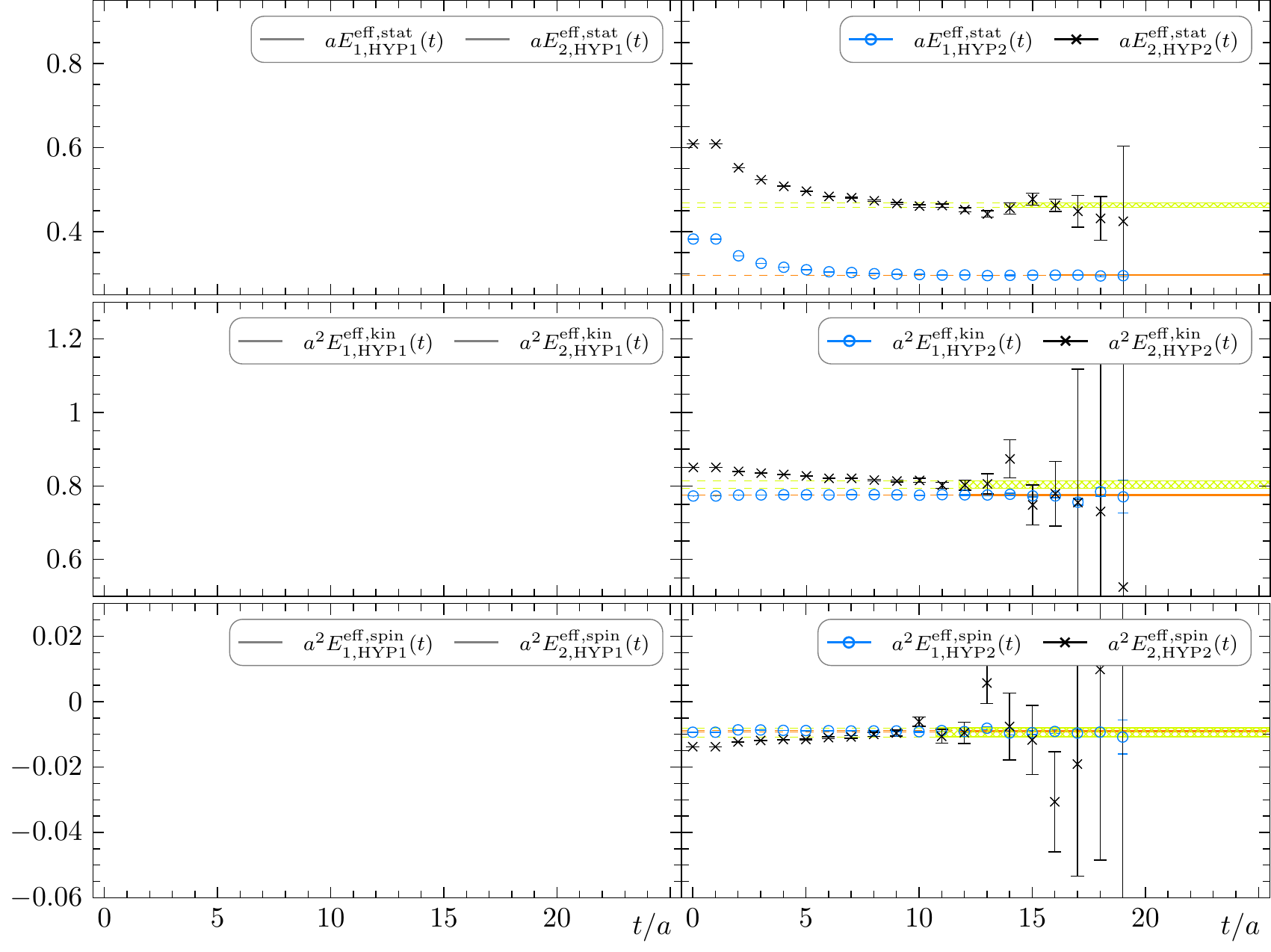}
   \caption{Effective energies $\Exnd$ following our GEVP analysis for 
            heavy-light (\emph{top}) and heavy-strange 
            (\emph{bottom}) sector on ensemble O7c.
           }
   \label{fig:Ex_O7c}
\end{figure}

%% file: tables/table_Eeff0_hl.tex
%
%
\begin{tabular}{ll@{\;\,}ll@{\;\,}ll@{\;\,}ll} \toprule
        & \multicolumn{2}{c}{$ a  E^{\rm stat}_{n=1}$} 
        & \multicolumn{2}{c}{$ a^2E^{\rm kin }_{n=1}$}   
        & \multicolumn{2}{c}{$-a^2E^{\rm spin}_{n=1}$}   
        \\ 
       \cmidrule(r){2-3}\cmidrule(r){4-5}\cmidrule(r){6-7}
 $e$-id& HYP1              & HYP2               & HYP1              & HYP2              & HYP1               & HYP2                \\ \midrule
    A4 & $0.4193(5)_{11} $ & $0.3841(4)_{11} $  & $0.8538(9)_{8}  $ & $0.8987(6)_{8}  $ & $0.02281(16)_{7} $ & $0.02183(13)_{7} $  \\
    A5c& $0.4151(5)_{10} $ & $0.3802(4)_{10} $  & $0.8520(6)_{7}  $ & $0.8962(4)_{7}  $ & $0.02253(18)_{7} $ & $0.02154(14)_{7} $  \\
    A5d& $0.4143(6)_{10} $ & $0.3791(5)_{10} $  & $0.8510(9)_{7}  $ & $0.8961(6)_{7}  $ & $0.02210(24)_{7} $ & $0.02105(20)_{7} $  \\
    B6 & $0.4069(10)_{11}$ & $0.3716(8)_{11} $  & $0.8490(10)_{7} $ & $0.8946(7)_{7}  $ & $0.02187(29)_{7} $ & $0.02067(22)_{7} $  \\[0.15em]
    E5 & $0.3845(4)_{12} $ & $0.3511(4)_{12} $  & $0.8115(9)_{9}  $ & $0.8523(6)_{9}  $ & $0.01762(19)_{9} $ & $0.01675(15)_{9} $  \\
    F6 & $0.3727(10)_{13}$ & $0.3392(8)_{13} $  & $0.8089(15)_{9} $ & $0.8490(9)_{9}  $ & $0.01697(33)_{9} $ & $0.01632(24)_{9} $  \\
    F7 & $0.3700(11)_{13}$ & $0.3354(9)_{13} $  & $0.8087(15)_{9} $ & $0.8486(9)_{9}  $ & $0.01563(33)_{9} $ & $0.01519(24)_{9} $  \\
    G8 & $0.3672(13)_{11}$ & $0.3327(11)_{11}$  & $0.8044(27)_{9} $ & $0.8470(16)_{9} $ & $0.01633(57)_{9} $ & $0.01508(38)_{9} $  \\[0.15em]
    N5 & $0.3195(5)_{16} $ & $0.2885(5)_{16} $  & $0.7414(15)_{12}$ & $0.7749(9)_{12} $ & $0.01018(25)_{12}$ & $0.00973(18)_{12}$  \\
    N6 & $0.3129(9)_{16} $ & $0.2825(6)_{16} $  & $0.7423(19)_{12}$ & $0.7727(11)_{12}$ & $0.00989(33)_{12}$ & $0.00899(21)_{12}$  \\
    O7 & $0.3093(10)_{16}$ & $0.2785(8)_{16} $  & $0.7389(26)_{12}$ & $0.7746(14)_{12}$ & $0.01000(42)_{12}$ & $0.00916(28)_{12}$  \\
    \bottomrule
\end{tabular}

%% file: tables/table_Eeff1_hl.tex
%
%
\begin{tabular}{ll@{\;\,}ll@{\;\,}ll@{\;\,}ll} \toprule
        & \multicolumn{2}{c}{$ a  E^{\rm stat}_{n=2}$} 
        & \multicolumn{2}{c}{$ a^2E^{\rm kin }_{n=2}$}   
        & \multicolumn{2}{c}{$-a^2E^{\rm spin}_{n=2}$}   
        \\ 
       \cmidrule(r){2-3}\cmidrule(r){4-5}\cmidrule(r){6-7}
 $e$-id& HYP1             & HYP2             & HYP1             & HYP2             & HYP1              & HYP2               \\ \midrule
    A4 & $0.655(9)_{10} $ & $0.622(8)_{10} $ & $0.931(22)_{9} $ & $0.975(16)_{9} $ & $0.0249(14)_{7} $ & $0.0210(12)_{7} $  \\
   A5c & $0.668(5)_{9}  $ & $0.633(4)_{9}  $ & $0.926(10)_{8} $ & $0.969(7)_{8}  $ & $0.0246(7)_{6}  $ & $0.0213(6)_{6}  $  \\
   A5d & $0.668(8)_{9}  $ & $0.631(7)_{9}  $ & $0.932(16)_{8} $ & $0.977(11)_{8} $ & $0.0268(11)_{6} $ & $0.0234(9)_{6}  $  \\
    B6 & $0.617(17)_{10}$ & $0.597(16)_{10}$ & $0.926(20)_{8} $ & $0.970(14)_{8} $ & $0.0256(14)_{6} $ & $0.0218(11)_{6} $  \\[0.15em]
    E5 & $0.608(7)_{11} $ & $0.577(6)_{11} $ & $0.878(10)_{9} $ & $0.915(6)_{9}  $ & $0.0204(13)_{8} $ & $0.0182(10)_{8} $  \\
    F6 & $0.590(13)_{11}$ & $0.556(12)_{11}$ & $0.862(17)_{9} $ & $0.901(11)_{9} $ & $0.0192(23)_{8} $ & $0.0157(18)_{8} $  \\
    F7 & $0.594(12)_{11}$ & $0.559(11)_{11}$ & $0.859(15)_{9} $ & $0.902(10)_{9} $ & $0.0072(67)_{10}$ & $0.0104(49)_{10}$  \\
    G8 & $0.577(27)_{11}$ & $0.548(23)_{11}$ & $0.853(13)_{8} $ & $0.906(9)_{8}  $ & $0.0277(36)_{8} $ & $0.0230(28)_{8} $  \\[0.15em]
    N5 & $0.490(5)_{13} $ & $0.462(4)_{13} $ & $0.805(30)_{13}$ & $0.779(17)_{13}$ & $0.0112(11)_{10}$ & $0.0103(8)_{10} $  \\
    N6 & $0.468(14)_{15}$ & $0.446(11)_{15}$ & $0.815(19)_{12}$ & $0.822(12)_{12}$ & $0.0113(14)_{10}$ & $0.0116(10)_{10}$  \\
    O7 & $0.465(17)_{14}$ & $0.432(13)_{14}$ & $0.777(8)_{10} $ & $0.815(5)_{10} $ & $0.0118(17)_{10}$ & $0.0096(13)_{10}$  \\
    \bottomrule
\end{tabular}

%% file: tables/table_Eeff0_hs.tex
%
%
\begin{tabular}{ll@{\;\,}ll@{\;\,}ll@{\;\,}ll} \toprule
        & \multicolumn{2}{c}{$ a  E^{\rm stat}_{n=1}$} 
        & \multicolumn{2}{c}{$ a^2E^{\rm kin }_{n=1}$}   
        & \multicolumn{2}{c}{$-a^2E^{\rm spin}_{n=1}$}   
        \\ 
       \cmidrule(r){2-3}\cmidrule(r){4-5}\cmidrule(r){6-7}
 $e$-id& HYP1               & HYP2               & HYP1              & HYP2             & HYP1               & HYP2                 \\ \midrule
    A4 & $0.43905(46)_{10}$ & $0.40430(40)_{10}$ & $0.8585(8)_{7}  $ & $0.9024(5)_{7}$  & $0.02365(22)_{7} $ & $0.02220(17)_{7} $   \\
    A5c& $0.43856(53)_{9} $ & $0.40378(46)_{9} $ & $0.8590(11)_{7} $ & $0.9006(7)_{7}$  & $0.02308(29)_{7} $ & $0.02214(24)_{7} $   \\
    B6 & $0.43224(41)_{12}$ & $0.39722(32)_{12}$ & $0.8556(4)_{7}  $ & $0.8990(3)_{7}$  & $0.02273(11)_{7} $ & $0.02139(9)_{7}  $   \\[0.15em]
    E5 & $0.40063(44)_{12}$ & $0.36736(36)_{12}$ & $0.8157(11)_{9} $ & $0.8561(7)_{9}$  & $0.01820(31)_{10}$ & $0.01753(21)_{10}$   \\
    F6 & $0.39468(35)_{14}$ & $0.36063(27)_{14}$ & $0.8122(6)_{9}  $ & $0.8530(4)_{9}$  & $0.01683(12)_{9} $ & $0.01612(9)_{9}  $   \\
    F7 & $0.39217(66)_{14}$ & $0.35837(54)_{14}$ & $0.8118(11)_{9} $ & $0.8534(7)_{9}$  & $0.01627(22)_{9} $ & $0.01576(16)_{9} $   \\[0.15em]
    N6 & $0.32827(44)_{17}$ & $0.29799(33)_{17}$ & $0.7429(11)_{12}$ & $0.7758(6)_{12}$ & $0.00991(18)_{12}$ & $0.00941(12)_{12}$   \\
    O7 & --                 & $0.29571(34)_{16}$ & --                & $0.7753(8)_{12}$ & --                 & $0.00905(15)_{12}$   \\
    \bottomrule
\end{tabular}

%% file: tables/table_Eeff1_hs.tex
%
%
\begin{tabular}{ll@{\;\,}ll@{\;\,}ll@{\;\,}ll} \toprule
        & \multicolumn{2}{c}{$ a  E^{\rm stat}_{n=1}$} 
        & \multicolumn{2}{c}{$ a^2E^{\rm kin }_{n=1}$}   
        & \multicolumn{2}{c}{$-a^2E^{\rm spin}_{n=1}$}   
        \\ 
       \cmidrule(r){2-3}\cmidrule(r){4-5}\cmidrule(r){6-7}
 $e$-id& HYP1              & HYP2              & HYP1             & HYP2             & HYP1              & HYP2               \\ \midrule
    A4 & $0.6829(77)_{9} $ & $0.6519(69)_{9} $ & $0.995(35)_{9} $ & $0.999(23)_{9} $ & $0.0250(12)_{6} $ & $0.0217(10)_{6} $  \\
    A5c& $0.6936(63)_{8} $ & $0.6614(58)_{8} $ & $0.952(20)_{8} $ & $0.975(15)_{8} $ & $0.0243(16)_{6} $ & $0.0212(13)_{6} $  \\
    B6 & $0.6762(61)_{10}$ & $0.6411(54)_{10}$ & $0.934(33)_{10}$ & $0.960(21)_{10}$ & $0.0224(11)_{7} $ & $0.0201(9)_{7}  $  \\[0.15em]
    E5 & $0.6022(82)_{11}$ & $0.5740(73)_{11}$ & $0.959(47)_{11}$ & $0.894(30)_{11}$ & $0.0170(17)_{8} $ & $0.0162(13)_{8} $  \\
    F6 & $0.6074(42)_{11}$ & $0.5730(37)_{11}$ & $0.855(13)_{10}$ & $0.914(8)_{10} $ & $0.0199(26)_{10}$ & $0.0147(18)_{10}$  \\
    F7 & $0.5885(68)_{11}$ & $0.5573(62)_{11}$ & $0.873(22)_{10}$ & $0.902(13)_{10}$ & $0.0186(16)_{8} $ & $0.0167(12)_{8} $  \\[0.15em]
    N6 & $0.4803(75)_{15}$ & $0.4554(58)_{15}$ & $0.781(12)_{12}$ & $0.813(7)_{12} $ & $0.0107(8)_{10} $ & $0.0105(6)_{10} $  \\
    O7 & --                & $0.4594(49)_{14}$ & --               & $0.807(9)_{12} $ & --                & $0.0091(12)_{11}$  \\
    \bottomrule
\end{tabular}

%% file: tables/table_peff0_hl.tex
%
%
\begin{tabular}{llllllll} \toprule
        & \multicolumn{2}{c}{$ a^{3/2}p^{\rm stat}_{n=1}$} 
        & \multicolumn{2}{c}{$-ap^{\rm kin }_{n=1}$}   
        \\ 
       \cmidrule(r){2-3}\cmidrule(r){4-5}
 $e$-id& HYP1              & HYP2              & HYP1             & HYP2              \\ \midrule
    A4 & $0.1223(5)_{10} $ & $0.1036(4)_{10} $ & $1.482(37)_{8} $ & $0.727(35)_{8} $  \\
   A5c & $0.1205(4)_{9}  $ & $0.1018(3)_{9}  $ & $1.501(19)_{6} $ & $0.748(19)_{6} $  \\
   A5d & $0.1192(6)_{9}  $ & $0.1012(5)_{9}  $ & $1.503(26)_{6} $ & $0.749(26)_{6} $  \\
    B6 & $0.1144(13)_{11}$ & $0.0970(10)_{11}$ & $1.473(44)_{7} $ & $0.743(41)_{7} $  \\[0.15em]
    E5 & $0.0984(4)_{11} $ & $0.0849(3)_{11} $ & $1.360(30)_{8} $ & $0.703(30)_{8} $  \\
    F6 & $0.0918(10)_{12}$ & $0.0787(8)_{12} $ & $1.329(49)_{8} $ & $0.668(45)_{8} $  \\
    F7 & $0.0896(9)_{12} $ & $0.0776(7)_{12} $ & $1.299(52)_{8} $ & $0.654(48)_{8} $  \\
    G8 & $0.0870(13)_{11}$ & $0.0755(10)_{11}$ & $1.370(56)_{7} $ & $0.703(52)_{7} $  \\[0.15em]
    N5 & $0.0620(6)_{15} $ & $0.0555(5)_{15} $ & $1.126(40)_{9} $ & $0.615(37)_{9} $  \\
    N6 & $0.0580(11)_{16}$ & $0.0510(7)_{16} $ & $1.145(69)_{10}$ & $0.599(61)_{10}$  \\
    O7 & $0.0559(13)_{16}$ & $0.0490(9)_{16} $ & $1.134(62)_{9} $ & $0.617(56)_{9} $  \\\midrule
        & \multicolumn{2}{c}{$ ap^{\rm A_0^{(1)}}_{n=1}$} 
        & \multicolumn{2}{c}{$ ap^{\rm spin}_{n=1}$}   
        \\ 
       \cmidrule(r){2-3}\cmidrule(r){4-5}
 $e$-id& HYP1              & HYP2              & HYP1              & HYP2              \\ \midrule
    A4 & $0.4451(9)_{11} $ & $0.3970(7)_{11} $ & $0.4792(5)_{5} $  & $0.4504(5)_{5} $  \\
   A5c & $0.4450(5)_{9}  $ & $0.3975(4)_{9}  $ & $0.4637(2)_{3} $  & $0.4385(2)_{3} $  \\
   A5d & $0.4440(14)_{9} $ & $0.3952(6)_{9}  $ & $0.4639(3)_{3} $  & $0.4385(3)_{3} $  \\
    B6 & $0.4422(13)_{10}$ & $0.3951(10)_{10}$ & $0.4793(10)_{5}$  & $0.4505(9)_{5} $  \\[0.15em]
    E5 & $0.4149(9)_{12} $ & $0.3677(6)_{12} $ & $0.4663(7)_{6} $  & $0.4393(7)_{6} $  \\
    F6 & $0.4121(17)_{12}$ & $0.3664(12)_{12}$ & $0.4673(11)_{6}$  & $0.4414(10)_{6}$  \\
    F7 & $0.4123(18)_{12}$ & $0.3645(12)_{12}$ & $0.4680(11)_{6}$  & $0.4426(11)_{6}$  \\
    G8 & $0.4111(21)_{11}$ & $0.3629(16)_{11}$ & $0.4698(19)_{6}$  & $0.4412(18)_{6}$  \\[0.15em]
    N5 & $0.3674(12)_{14}$ & $0.3207(8)_{14} $ & $0.4361(14)_{8}$  & $0.4149(13)_{8}$  \\
    N6 & $0.3583(20)_{15}$ & $0.3171(13)_{15}$ & $0.4351(16)_{8}$  & $0.4169(14)_{8}$  \\
    O7 & $0.3645(21)_{14}$ & $0.3200(13)_{14}$ & $0.4362(22)_{8}$  & $0.4139(18)_{8}$  \\
    \bottomrule
\end{tabular}

%% file: tables/table_peff0_hs.tex
%
%
\begin{tabular}{llllllll} \toprule
        & \multicolumn{2}{c}{$ a^{3/2}p^{\rm stat}_{n=1}$} 
        & \multicolumn{2}{c}{$-ap^{\rm kin }_{n=1}$}   
        \\ 
       \cmidrule(r){2-3}\cmidrule(r){4-5}
 $e$-id& HYP1               & HYP2               & HYP1             & HYP2             \\ \midrule
    A4 & $0.13448(77)_{10}$ & $0.11329(58)_{10}$ & $1.553(37)_{7} $ & $0.757(36)_{7} $ \\
    A5 & $0.13457(78)_{9} $ & $0.11269(60)_{9} $ & $1.563(34)_{6} $ & $0.781(33)_{6} $ \\
    B6 & $0.13143(50)_{11}$ & $0.11041(37)_{11}$ & $1.536(31)_{8} $ & $0.747(29)_{8} $ \\[0.15em]
    E5 & $0.10746(71)_{12}$ & $0.09264(51)_{12}$ & $1.391(41)_{8} $ & $0.712(39)_{8} $ \\
    F6 & $0.10521(47)_{13}$ & $0.09019(34)_{13}$ & $1.387(35)_{9} $ & $0.689(31)_{9} $ \\
    F7 & $0.10296(82)_{13}$ & $0.08857(61)_{13}$ & $1.403(38)_{8} $ & $0.703(38)_{8} $ \\[0.15em]
    N6 & $0.06562(56)_{16}$ & $0.05788(40)_{16}$ & $1.173(46)_{10}$ & $0.611(41)_{10}$ \\
    O7 & --                 & $0.05745(48)_{15}$ & --               & $0.620(39)_{9} $ \\\midrule
        & \multicolumn{2}{c}{$ ap^{\rm A_0^{(1)}}_{n=1}$} 
        & \multicolumn{2}{c}{$ ap^{\rm spin}_{n=1}$}   
        \\ 
       \cmidrule(r){2-3}\cmidrule(r){4-5}
 $e$-id& HYP1              & HYP2               & HYP1             & HYP2             \\ \midrule
    A4 & $0.4426(9)_{10} $ & $0.39451(72)_{10}$ & $0.4778(9)_{5} $ & $0.4488(9)_{5} $ \\
   A5c & $0.4395(10)_{9} $ & $0.39354(72)_{9} $ & $0.4782(13)_{5}$ & $0.4493(12)_{5}$ \\
    B6 & $0.4374(8)_{12} $ & $0.39070(61)_{12}$ & $0.4799(5)_{5} $ & $0.4505(5)_{5} $ \\[0.15em]
    E5 & $0.4164(10)_{12}$ & $0.36767(78)_{12}$ & $0.4657(10)_{6}$ & $0.4393(9)_{6} $ \\
    F6 & $0.4117(6)_{12} $ & $0.36459(42)_{12}$ & $0.4658(6)_{6} $ & $0.4398(5)_{6} $ \\
    F7 & $0.4094(12)_{13}$ & $0.36180(93)_{13}$ & $0.4670(10)_{6}$ & $0.4395(8)_{6} $ \\[0.15em]
    N6 & $0.3605(13)_{16}$ & $0.31679(90)_{16}$ & $0.4364(10)_{8}$ & $0.4165(9)_{8} $ \\
    O7 & --                & $0.31636(88)_{15}$ & --               & $0.4176(13)_{8}$ \\
    \bottomrule
\end{tabular}